\documentclass[aps,a4paper,twocolumn]{revtex4-2}
\usepackage{amsmath}
\usepackage{graphicx}
\usepackage{subfigure}
\usepackage[usenames,dvipsnames]{color}
\definecolor{darkblue}{RGB}{0,0,196}
\usepackage[colorlinks=true,linkcolor=darkblue,citecolor=darkblue,urlcolor=darkblue]{hyperref}
\usepackage{setspace}
\usepackage{hyperref}
\usepackage{xcolor}
\hypersetup{
  colorlinks   = true, 
  urlcolor     = red, 
  linkcolor    = blue, 
  citecolor   = blue 
}
\usepackage{graphicx}
\usepackage{amsmath,bbm}
\usepackage{amssymb,bm}
\usepackage{yfonts}
\usepackage{comment}

\newcommand{\dca}{DCA$_{D^{0}}$}
\usepackage[usenames,dvipsnames]{color}
\definecolor{darkblue}{RGB}{0,0,196}
\usepackage[colorlinks=true,linkcolor=darkblue,citecolor=darkblue,urlcolor=darkblue]{hyperref}

\begin{document}
\title{A machine learning-based study of open-charm hadrons in proton-proton collisions at the Large Hadron Collider}

\author{Kangkan Goswami} 
\author{Suraj Prasad}
\author{Neelkamal Mallick} 
\author{Raghunath Sahoo}
\email{Corresponding Author: Raghunath.Sahoo@cern.ch}
\affiliation{Department of Physics, Indian Institute of Technology Indore, Simrol, Indore 453552, India}
\author{Gagan B. Mohanty}
\affiliation{Tata Institute of Fundamental Research, Homi Bhabha Road, Mumbai 400005, India}

\begin{abstract}
In proton-proton and heavy-ion collisions, the study of charm hadrons plays a pivotal role in understanding the QCD medium and provides an undisputed testing ground for the theory of strong interaction, as they are mostly produced in the early stages of collisions via hard partonic interactions. The lightest open-charm, $D^{0}$ meson ($c\Bar{u}$), can originate from two separate sources. The prompt $D^{0}$ originates from either direct charm production or the decay of excited open charm states, while the nonprompt stems from the decay of beauty hadrons. In this paper, using different machine learning (ML) algorithms such as XGBoost, CatBoost, and Random Forest, an attempt has been made to segregate the prompt and nonprompt production modes of $D^{0}$ meson signal from its background. The ML models are trained using the invariant mass through its hadronic decay channel, i.e., $D^{0}\rightarrow\pi^{+} K^{-}$, pseudoproper time, pseudoproper decay length, and distance of closest approach of $D^{0}$ meson, using PYTHIA8 simulated $pp$ collisions at $\sqrt{s}=13~\rm{TeV}$. The ML models used in this analysis are found to retain the pseudorapidity, transverse momentum, and collision energy dependence. In addition, we report the ratio of nonprompt to prompt $D^{0}$  yield, the self-normalized yield of prompt and nonprompt $D^{0}$ and explore the charmonium, $J/\psi$ to open-charm, $D^{0}$ yield ratio as a function of transverse momenta and normalized multiplicity. The observables studied in this manuscript are well predicted by all the ML models compared to the simulation.
\end{abstract}
\maketitle
\section{Introduction}
To understand the fundamental nature of our Universe, accelerator facilities such as the Relativistic Heavy-Ion Collider (RHIC) at BNL and the Large Hadron Collider (LHC) at CERN perform proton-proton ($pp$) and heavy ion collisions at ultra-relativistic speeds~\cite{STAR:2005gfr, ALICE:2022wpn}. These collisions allow us to explore a unique state of thermalized and deconfined medium of quarks and gluons, known as the quark-gluon plasma (QGP)~\cite{Cleymans:1985wb, Yagi:2005yb, Sarkar:2010zza, Gyulassy:2004zy}. Understanding the QGP medium, which mimics conditions of the micro-second-old Universe, is crucial. Furthermore, the study of the created matter at the extreme conditions of temperatures and energy densities sheds light on the phase transition from the deconfined partonic phase to the color-neutral hadronic phase, where they are confined within the hadrons, thereby making a testing ground for QCD strong interaction dynamics. However, the QGP medium is extremely transient, having a lifetime of the order of $10^{-23}~\rm{s}$, before the quarks and gluons hadronize into a color-neutral state~\cite{Yagi:2005yb, Sarkar:2010zza}. As a result, we can only detect the final-state hadrons after the kinetic freeze-out. Therefore, precise probes are essential to investigate the characteristics of this deconfined partonic medium.

One such probe for the study of the deconfined phase is the heavy quarks (HQs), i.e., charm and beauty. The HQs are produced in the initial hard scattering. Their production time is characterized by $\Delta t > (\frac{1}{2m_{c,b}})$; $\sim 0.1~\rm{fm/c}$ for charm quarks and $\sim 0.01~\rm{fm/c}$ for beauty quarks, which is much shorter than the formation time ($\sim 0.3~\rm{fm/c}$) of the QGP medium~\cite{Liu:2012ax, ALICE:2015vxz}. In addition, due to their masses being much larger than the temperature of the QGP medium, the probability of thermal production and annihilation of HQs is negligible. The HQs undergo Brownian motion in the thermalized medium of lighter quarks ($u, d, s$) and experience the entire evolution of the QGP medium. These HQs combine with the light-flavor quarks at the phase boundary or during the system evolution to form the open-heavy-flavor hadrons. The most abundant of them is the $D^{0}$ meson ($c\Bar{u}$) due to its lowest mass among all the heavy-flavor hadrons. The $D^{0}$ meson originates from two sources following different topologies. First,  the prompt $D^{0}$ mesons come directly from the initial hard scatterings as the feed down from the higher excited charm states ($D^{*}~(\rm{2007})^{0}$, $D_{1}(\rm{2420})^{+}$). Second, the nonprompt $D^{0}$ meson comprises charm quarks that are produced through flavor-changing weak decays of beauty hadrons~\cite{CDF:2003vmf, ALICE:2018lyv}. It is essential to separate the prompt and nonprompt $D^{0}$ to understand the relative contribution from the charm and beauty sectors. This helps in studying the nuclear modification factor in the charm and beauty sectors separately, which may shed light on possible different mechanisms of energy loss in the QCD medium~\cite{ALICE:2022tji}. Further, this facilitates the study of different phenomena like HQ transport and thermalization in the medium through anisotropic flow~\cite{ALICE:2023gjj}.
In addition, the study of topological production of $D^0$ meson has several physics implications. Prompt $D^{0}$ meson can help to understand the QCD medium and can provide a testing ground for the theory of strong interactions. Various observables are measured in experiments from the final-state hadrons to understand the interaction of the charm quarks in the QGP medium, where a comprehensible insight can be gained using prompt $D^0$ mesons as a probe. Since the nonprompt $D^0$ mesons are weak decay products of beauty hadrons, they are produced at a larger distance from the primary interaction vertex. Thus, using non-prompt $D^0$ mesons to understand initial partonic interactions may not be an ideal choice. On the other hand, the nonprompt production of the $D^{0}$ mesons can help to unveil the beauty production in both $pp$ and heavy-ion collision sectors~\cite{ALICE:2022tji, ALICE:2024xln}.

Experimentally, the study of heavy-flavor (charm and beauty) hadrons acts as a good testing ground for perturbative Quantum Chromodynamics (pQCD) calculations. The topological separation of prompt and nonprompt $D$ meson is an important aspect of studying the production and evolution of charm and beauty quarks in the QCD medium~\cite{ALICE:2022xrg}. Moreover, the azimuthal anisotropy in the momentum space of final-state hadrons serves as an excellent probe of the QGP medium. The second order anisotropic flow coefficient, known as the elliptic flow or $v_2$, has been calculated for $D$ mesons in ALICE, STAR, and CMS experiments ~\cite{ALICE:2013olq, ALICE:2021kfc, STAR:2017kkh, CMS:2018loe, CMS:2020bnz}. Recently, the elliptic flow of the nonprompt $D^{0}$ meson has been estimated for Pb--Pb collisions at CMS and ALICE ~\cite{CMS:2022vfn, ALICE:2023gjj}. This helps in studying the degree of charm and beauty quark thermalization and their participation in the collective expansion of the medium. Additionally, the nuclear modification factor ($R_{\rm AA}$) is estimated to explore the energy loss by the HQs through interaction with the medium, taking $pp$ collisions as a baseline~\cite{CMS:2017qjw, ALICE:2015vxz, ALICE:2021rxa}. The advancements in Run-3 detector upgrades and the high luminosity of ALICE offer a significant opportunity for the thorough and rigorous exploration of the charm and beauty hadron production in hadronic and nuclear collisions.

\begin{figure}
    \centering
    \includegraphics[width = \linewidth]{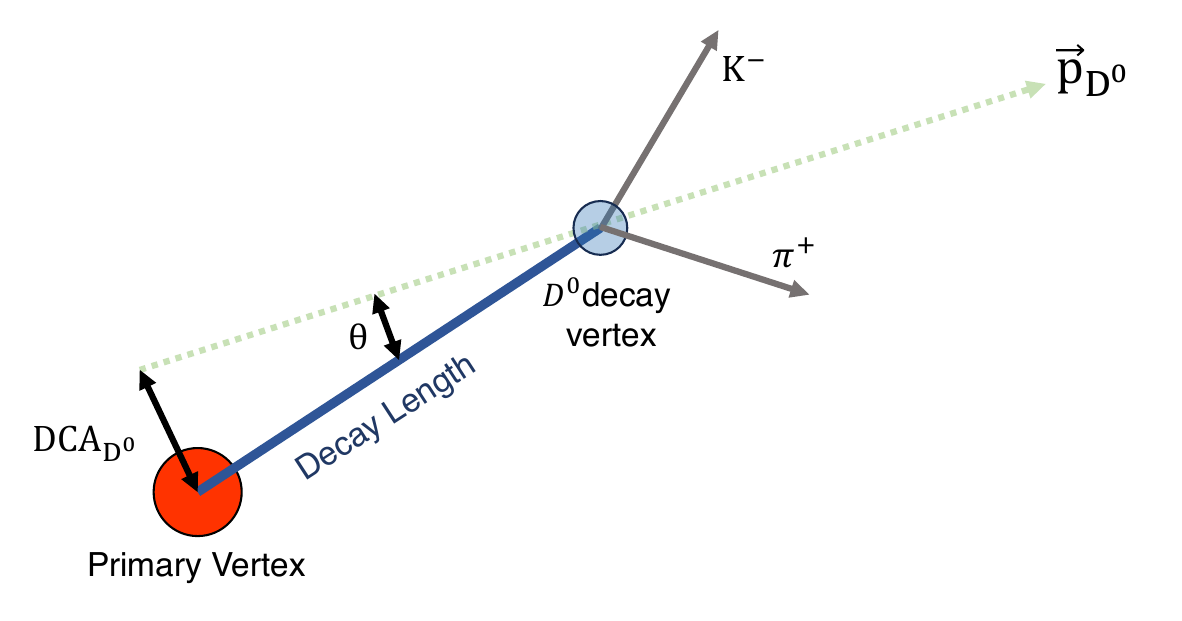}
    \caption{Schematic diagram of $D^{0}$ meson production and decay topology in hadronic and nuclear collisions.}
     \label{diag}
\end{figure}

Typically, $D^{0}$ meson is reconstructed through its hadronic decay channel $D^{0} \rightarrow \pi^{+} K^{-}$. The inclusive $D^{0}$ is dominated by prompt $D^{0}$ contributions, with only a small fraction being nonprompt $D^{0}$. Figure~\ref{diag} provides a schematic diagram of $D^{0}$ production and  $D^{0}\rightarrow \pi^{+} K^{-}$ decay topology. The decay length represents the distance between the $D^{0}$ decay vertex and the primary vertex. The distance of the closest approach of the $D^{0}$ meson (\dca) is measured by taking the distance between the primary vertex and the reconstructed $D^{0}$ momentum vector $\vec{p}_{D^{0}}$. The beauty hadrons undergo a weak decay into a $D^{0}$ meson, which further decays into a $\pi^{+} K^{-}$ pair, whereas the prompt $D^{0}$ mesons are produced much closer to the primary vertex. The involvement of the weak interaction in the decay topology of the nonprompt $D^{0}$ meson increases the distance between the primary vertex and $D^{0}$ decay vertex. Consequently, the \dca for the nonprompt $D^{0}$ mesons is higher than the prompt counterparts. 

In this study, we take advantage of the machine learning (ML) techniques to separate the contribution from the charm and beauty sector by classifying the prompt and nonprompt $D^{0}$ mesons using final-state observables as input features. The ML algorithms, with proper training, can map a correlation between the input features and output. This is achieved through building a classification model from sample inputs, which allows the machine to learn independently and build a correlation between the inputs and outputs. The ML algorithms are categorized into supervised, unsupervised, semi-supervised, and reinforcement learning, each having its unique approach and application. In the case of experimental high-energy physics, the potential of machine learning lies in its ability to discover correlations in large datasets. The ML techniques have been in use in the field of high-energy for the last few decades~\cite{Bowser-Chao:1992giy, Chiappetta:1993zv, Bass:1996ez}. It is successfully deployed for studies like jet mesurements~\cite{Haake:2019pqd, ALICE:2023waz, Paganini:2017dpd, ATLAS:2020iwa}, particle identification~\cite{CMS:2020poo, Graczykowski:2022zae, Ryzhikov:2022lbu}, impact parameter estimation~\cite{Bass:1996ez, Mallick:2021wop, Zhang:2021zxd}, flow coefficient measurements~\cite{Mallick:2022alr, Mallick:2023vgi, Hirvonen:2023lqy}. Recently, classification problems, such as classifying prompt and nonprompt $J/\psi$ in forward rapidity~\cite{Prasad:2023zdd} and segregating electrons coming from different sources~\cite{ATLAS:2023mnn} were successfully addressed. For our study of prompt and nonprompt classification of $D^{0}$, we simulate $pp$ collisions at $\sqrt{s} = 13~\rm{TeV}$ using PYTHIA8 and train three different ML algorithms, namely XGBoost, CatBoost, and Random Forest. On successful training, we use our ML models to predict the production of prompt and nonprompt $D^{0}$ mesons for $pp$ collisions at $\sqrt{s}=5.02~\rm{TeV}$ and $900~\rm{GeV}$. The novelty of the work is reflected in the model's robustness in distinguishing between prompt and nonprompt $D^{0}$ particles throughout the entire energy range of the LHC. On separating the prompt and nonprompt $D^{0}$ mesons, we attempt to understand their production dynamics with respect to the production of charged particles and charmonium state ($J/\psi$).

The remainder of the paper is organized as follows: in Section~\ref{method}, we briefly discuss the methodology of the work. Section~\ref{train_eval} starts with a discussion of the training and evaluation of the models. Followed by the results and discussion in Section~\ref{result_dis}. Finally, we summarize and conclude our findings in Section~\ref{summ}.

\section{Methodology}
\label{method}
In this section, we present a brief introduction to event generation using PYTHIA8, followed by ML algorithms. Additionally, the production cross-sections of prompt and nonprompt $D^0$ meson obtained from simulation are compared to published measurements from ALICE to quality check the tunes and settings used in PYTHIA8. 

\subsection{PYTHIA8}
\label{pythia8}
Event generators are used to simulate hadronic and heavy-ion collisions with greater control over the evolution stages and to test various phenomenological models. These generators use Monte Carlo simulation techniques to mimic the actual collisions involving a variety of physics processes. PYTHIA8, a Monte Carlo event generator, is commonly employed to simulate ultra-relativistic hadronic, leptonic, as well as heavy-ion collisions across a wide range of energy. It provides a comprehensive explanation of the pQCD-based particle production, including charm and beauty production. In this study, we use PYTHIA8 to simulate events for the training of ML algorithms to distinguish prompt and nonprompt 
particles.

\begin{figure}[!ht]
    \centering
    \includegraphics[width = \linewidth]{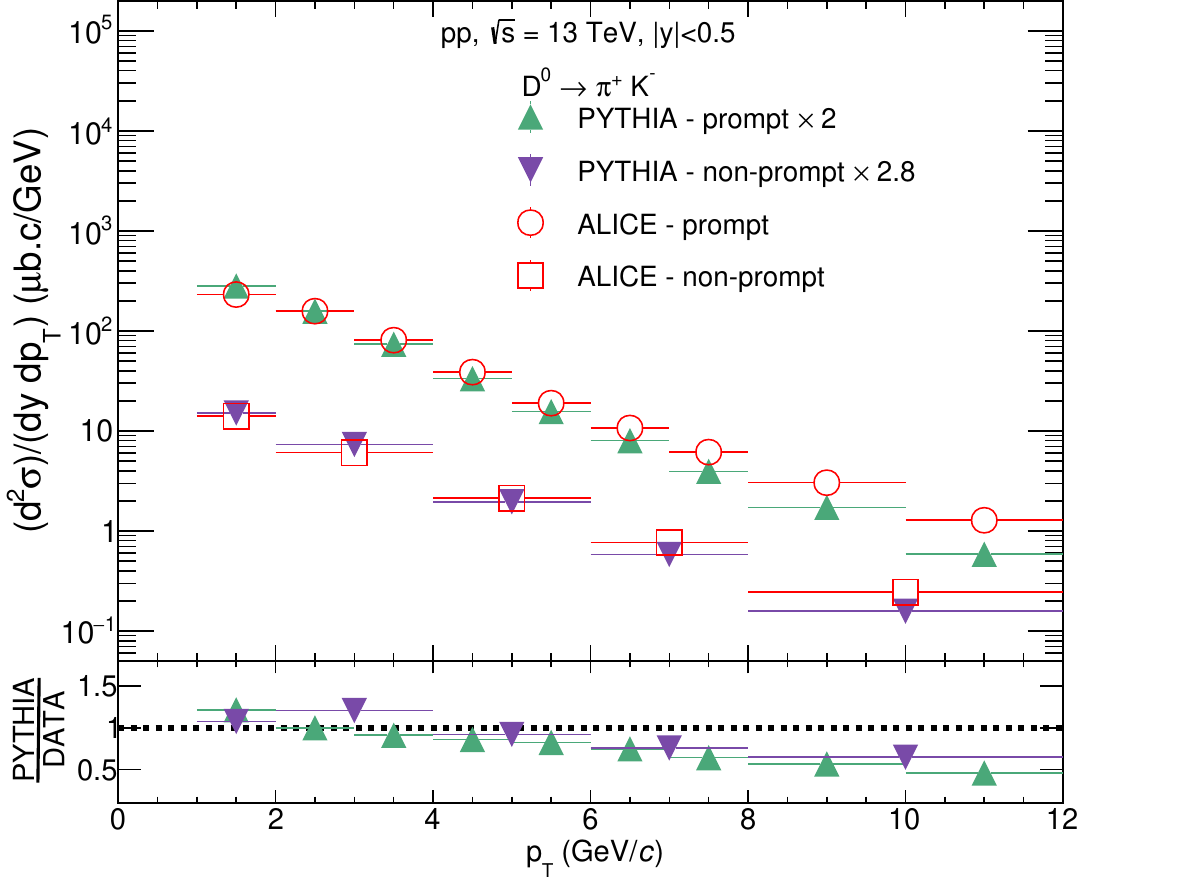}
    \caption{Upper panel shows the prompt and nonprompt $D^{0}$ meson production cross-section in $pp$ collisions at $\sqrt{s} = 13~\rm{TeV}$ generated with PYTHIA8, compared with ALICE data~\cite{ALICE:2021rzj, ALICE:2023wbx}. The lower panel depicts the PYTHIA8 to ALICE data ratio.}
    \label{fig:ALICE_pT_spectra}
\end{figure}

PYTHIA8 consists of particle production mechanisms involving soft and hard processes, initial and final state parton shower, string fragmentation, hadronic rescattering and decay, color reconnection, beam remnants, and multiple parton interactions (MPI). This is an improved version of PYTHIA6 that incorporates a scenario based on MPI. In this scenario, $2 \rightarrow 2$ hard processes have the potential to generate heavy quarks such as charm and beauty. For this study, we have used PYTHIA 8.308, with 4C tune~\cite{Corke:2010yf} ($\rm{tune:pp= 5}$) and considering only inelastic and non-diffractive components ($\rm{HardQCD:all=on}$), to generate 2 billion minimum bias events for $pp$ collisions at $\sqrt{s} = 13~\rm{TeV}$. Furthermore, we generate 1 billion minimum bias events for $pp$ collisions at $\sqrt{s} = 5.02~\rm{TeV}$ and $900~\rm{GeV}$ each. To prevent the divergence of QCD processes, which can happen when transverse momentum, $p_{\rm T} \rightarrow 0$, we implement a $p_{\rm T}$ cutoff of 0.5 GeV/c ($\rm{PhaseSpace:pTHatMinDiverge = 0.5}$). The data have been simulated with color reconnection taken into consideration ($\rm{ColourReconnection: reconnect = on}$). Additionally, we have utilized the mode-2 for color reconnection, indicated by $\rm{ColourReconnection: mode = 2}$. This mode refers to the gluon-move model, where the gluons are moved (or flipped) from one point to another such that the string length is minimized~\cite{PYTHIA8}.
For the production of prompt and nonprompt $D^{0}$ mesons, we have enabled all the charmonium and bottomonium production processes via $\rm{Charmonium:all=on}$ and $\rm{Bottomonium:all=on}$. A detailed description of the physics processes and their implementation in PYTHIA8 are provided in Ref.~\cite{Sjostrand:2006za, PYTHIA8}.

To mimic the real-world experiments, we enable the spread of the primary interaction vertex following a Gaussian distribution ($\rm{Beams:allowVertexSpread=on}$) as also done in Ref.~\cite{Prasad:2023zdd}. The mean and standard deviation of the distribution in the cartesian coordinate are taken from Ref.~\cite{ALICE:2010vtz}. Following the experimental methods, we have also taken a cutoff at the $z$ component of the interaction vertex, i.e., $|V_{z}|<10~\rm{cm}$. We have allowed the decay of $D^{0}$ through all the possible decay modes. In PYTHIA8, we examine the mother of the reconstructed $D^{0}$ meson to classify it into prompt or nonprompt $D^{0}$. In Fig.~\ref{fig:ALICE_pT_spectra}, we compare the PYTHIA8 generated $p_{\rm T}$ spectra with recent ALICE results~\cite{ALICE:2021rzj, ALICE:2023wbx}. It is noteworthy that the CMS and LHCb experiments have measured only prompt $D^{0}$ in $pp$ collisions, and their kinematic ranges are different \cite{CMS:2021lab,LHCb:2015swx}.
One can readily observe that the normalized yield of the prompt $D^{0}$ is around 10 times higher than the production of nonprompt $D^{0}$. The similar difference is continued up to the high-$p_{\rm T}$ range of around $12~\rm{GeV/c}$. PYTHIA8 underestimates the ALICE data, and hence, a factor of $2.0$ and $2.8$ is multiplied by the prompt and nonprompt yield, respectively to match the spectral shape. The trend of the $p_{\rm T}$ spectra shown by PYTHIA8, with all the above-mentioned tunes, is comparable with ALICE data as seen from the lower ratio plot. However, for the rest of the results, we do not apply any scaling factor to the PYTHIA8-generated spectra.

\subsection{Machine Learning Algorithms}

With the introduction of ML tools, drawing significant conclusions from a large set of experimental data has become easier and more reliable. This is achieved by properly taking care of the correlations among the input features. In experimental high-energy physics, one of the most complex problems is understanding the underlying physical processes in particle production in the subatomic realm. With the detected final state particles, one can use their four-momenta as input features to the ML algorithm. In this study, we use three ML algorithms, namely, CatBoost (v1.2), Random Forest (v1.3.0), and XGBoost (v1.7.3). These techniques are very efficient for classification problems, each with its unique strength.  

These three models are often preferred over others due to their robustness, efficiency, and the fact that they can easily handle a variety of data types. They also have the ability to model complex nonlinear correlations, which adds to their versatility and utility in many real-world applications~\cite{Torfi:2019}. For training and prediction, we use Python 3.11 as well as computing and plotting the confusion matrix, importance score, and learning curve. 

\subsubsection{CatBoost}
CatBoost (CB) stands for Categorical Boosting. It is a high-performance ML algorithm that has gained popularity due to its ability to handle categorical data directly, with no need for manual one-hot encoding~\cite{catboost, Prokhorenkova:2017}. It also implements ordered boosting, a permutation-driven alternative to the classical algorithm, which improves the model prediction. It is an implementation of gradient boosting designed to combat the problem of overfitting by implementing a novel algorithm for calculating leaf values. CatBoost also supports Graphics Processing Unit acceleration, which can significantly speed up the training process. It provides a wide range of hyper-parameters that can be fine-tuned to improve the model’s performance.

\subsubsection{Random Forest}
Random Forest (RF) is a versatile and widely used ML algorithm that operates by constructing multiple decision trees during training. It gives the output as the class, i.e., the mode of the classes for classification or mean prediction for regression tasks~\cite{Louppe:2014}. It is highly flexible and efficient, even without hyperparameter tuning. One of the key advantages of Random Forest is that it can be used for both regression and classification tasks. It provides a good indicator of the feature's importance, handles high-dimensional spaces well, and can deal with unbalanced datasets. Random Forest is also less likely to overfit than individual decision trees.

\subsubsection{XGBoost}
XGBoost (XGB), which stands for Extreme Gradient Boosting, is a highly regarded and extensively utilized ML algorithm~\cite{Chen:2016, readthedocs}. It is particularly effective in dealing with large datasets and excels in both classification and regression tasks. XGB is an advanced version of Gradient-Boosting Decision Trees (GBDT) and includes several improvements, such as parallel computing and tree pruning. These enhancements expedite the training process, enabling XGB to manage large datasets efficiently. Furthermore, XGB offers a broad range of hyperparameters that can be fine-tuned to enhance the performance of the model.

\section{Training and Evaluation}
\label{train_eval}

In this section, the topological features used as inputs for the ML models are defined, followed by data pre-processing and model training. Finally, a few quality assurance plots are presented to demonstrate the classification accuracy of the ML models.


\subsection{Input to the machine}
In this study, a few topological features are selected as the inputs to the ML models. Our goal is to utilize such features that can identify the topological production dynamics of prompt and nonprompt $D^{0}$ mesons. First, the inclusive $D^{0}$ meson signal has to be identified over the background, followed by the identification of the prompt and nonprompt production modes. One can identify the inclusive $D^{0}$ meson signal from the background with the help of its invariant mass ($m_{\pi K}$), where a peak in the $m_{\pi K}$ distribution is observed around the $D^{0}$ mass. The identification of prompt and nonprompt $D^{0}$ can then be performed by looking at the variables sensitive to their decay topology. For example, as the prompt $D^{0}$ mesons are produced closer to the primary vertex as compared to the nonprompt case, this eventually leads to a larger decay length for the $D^{0}$ mesons coming from the decay of beauty hadrons. The topological variables associated with the displaced production vertex of the $D^{0}$ mesons are the pseudoproper time ($t_z$)~\cite{LHCb:2017ygo}, the pseudoproper decay length ($c\tau$)~\cite{ ALICE:2021edd}, and the angle ($\theta$) between the $D^0$ momentum vector and the vector joining the $D^0$ decay vertex to the primary vertex~\cite{STAR:2018zdy}. The pseudoproper time is defined as~\cite{LHCb:2017ygo},  
\begin{equation}
    t_{z} = \frac{(z_{D^{0}} - z_{\rm PV})\times m_{D^{0}}}{p_{z}}
\end{equation}
where $z_{D^{0}}$ and $z_{\rm PV}$ are the coordinates of the $D^{0}$ decay vertex and primary vertex along the beam direction ($z$-axis), $m_{D^{0}} \simeq 1865$~MeV is the mass of the $D^{0}$ meson taken from the Particle Data Group~\cite{ParticleDataGroup:2022pth}, and $p_{z}$ is the momentum in the $z$-direction. The decay topology of the $D^0$ meson in the longitudinal direction is quantified by $t_{z}$, where $t_{z}$ is expected to have a higher value for the nonprompt $D^0$ mesons as compared to the prompt $D^0$ mesons that are produced closer to the primary vertex.

Similarly, one can also quantify the decay topology of the particles in the transverse plane using pseudoproper decay length ($c \tau$).
One can write the pseudoproper decay length as~\cite{ALICE:2021edd},
\begin{equation}
    c \tau = \frac{c ~m_{D^{0}}\Vec{L}.\Vec{p_{T}}}{|p_{T}|^{2}}
\end{equation}
where, $\Vec{L}$ is a vector pointing from the primary vertex towards the $D^{0}$ decay vertex, i.e. $\Vec{L} = \Vec{V} - \Vec{S}$. Here, $\Vec{V}=(V_{x}, V_{y}, V_{z})$ is the position of the primary vertex and $\Vec{S}=(S_{x}, S_{y}, S_{z})$ is the position of $D^{0}$ meson decay vertex with respect to the global origin, i.e., ($0,0,0$). As already mentioned in Sec.~\ref{pythia8}, we have used a Gaussian profile to randomize the position of the primary vertex in three dimensions to be consistent with an experimental scenario. In experiments, we can reconstruct the $D^0$ decay vertex as the middle point on the distance of the closest approach between the candidate pion and kaon trajectories. However, in PYTHIA8, this is not trivial, and therefore, we need to estimate the $D^0$ decay vertex ($\Vec{S}$). One can calculate the same by using the following expression~\cite{Prasad:2023zdd}.
\begin{equation}
    S_{i} = \frac{(t_{1} + d_{i,1}m_{1}/p_{i,1}) - (t_{2} + d_{i,2}m_{2}/p_{i,2})}{m_{1}/p_{i,1} - m_{2}/p_{i,2}}
\end{equation}
where $i = x,y,z$ is the spatial index, and $m_{1}$ and $m_{2}$ are the masses of the two decay products of the $D^{0}$ meson. $d_{i,1}$ and $d_{i,2}$ are the distances covered by the decay products in time $t_{1}$ and $t_{2}$ with momentum $p_{i,1}$ and $p_{i,2}$, respectively. Thus, using $\Vec{V}$, and $\Vec{S}$, one can obtain the value of $\Vec{L}$ and consequently estimate the value of $c\tau$.

Finally, we use \dca, which is well estimated in experiments, as another topological input variable to the ML models. \dca is defined in terms of the decay length and sine of the angle between $\Vec{L}$ and the $D^0$ momentum vector $\vec{p}_{D^{0}}$ as~\cite{STAR:2018zdy},
\begin{equation}
    {\rm DCA}_{D^0} = |\Vec{L}| \times \sin\theta.
\end{equation}
As discussed earlier, due to the difference in the production topology of prompt and nonprompt $D^{0}$ mesons, we can expect larger \dca for the nonprompt $D^{0}$ meson. Thus, we proceed to train the ML models with $m_{\pi K}$, and the above discussed topological variables such as $t_{z}$, $c\tau$ and \dca  ~of the reconstructed $\pi^{+}K^{-}$ pairs as the input variables to the machine. The training is performed using 600 million minimum bias $pp$ collisions at $\sqrt{s} = 13~\rm{TeV}$.

\subsection{Pre-processing and training}

The task of the ML models is to classify the prompt and nonprompt $D^{0}$ mesons from the background using the topological features of the reconstructed $\pi^{+}K^{-}$ pairs. However, the number of prompt $\pi^{+}K^{-}$ pairs is naturally smaller than that of uncorrelated background pairs. Similarly, the $\pi^{+}K^{-}$ pairs coming from nonprompt $D^{0}$ meson are even smaller as compared to the prompt pairs owing to the smaller production cross-section of charm quarks from beauty decays than the direct charm production as shown in Fig.~\ref{fig:ALICE_pT_spectra}. Hence, an ML model trained with such a dataset shows a bias towards the most populated class, in our case, the background class. Therefore, the trained model will show a higher degree of inaccuracy by frequently predicting the most populated class when applied to a testing set. This is known as the class imbalance problem. Thus, the pre-processing of the input dataset becomes essential to avoid this class imbalance problem, which also enhances the quality of the training data. This leads to unbiased training that improves the classification accuracy of the ML models.


The class imbalance problem is often addressed via sampling techniques. We could use different sampling techniques to pre-process our training data, such as the oversampling or undersampling methods. Undersampling involves reducing the number of samples from the majority class to balance the number of instances from each class in the dataset. However, this has a serious downside as it can discard potentially useful data during the process by reducing the training statistics. Oversampling, on the other hand, involves increasing the samples in the minority class. This is achieved by duplicating the samples in the minority class. However, creating duplicate copies of the data may sometimes lead to overfitting. In this study, we use the Synthetic Minority Over-sampling Technique (SMOTE) to create new samples for the minority classes~\cite{Chawla:2002dkc}. SMOTE creates synthetic samples from the minority class instead of creating copies. By doing this, SMOTE provides better information to the model about the minority class. Before oversampling, the ratio background:prompt:nonprompt was 50:20:1, which changed to 15:5:1 after oversampling using SMOTE. Moreover, for training, testing, and validation purposes, we split our input data into 8:1:1 (train:test:validation) set. To test the stability of SMOTE oversampling, we compared the results with
Random Oversampling (ROS) and weighted Random Oversampling (wROS) methods, and observed similar levels of prediction
accuracy.

With this pre-processed data and class imbalance challenge out of sight, we proceed to train the ML algorithms. The optimum hyperparameters related to the XGBoost, CatBoost, and Random Forest models are listed in Table \ref{Table1}, \ref{Table2}, and \ref{Table3}, respectively, and are briefly discussed in the next paragraph.

\begin{table}[h]
\centering
\textbf{XGBoost (XGB)} \\
\begin{tabular}{l c }
\hline
\hline
Parameter & Value \\
\hline
    booster & \textit{gbtree} \\
    learning$\_$rate & 0.3 \\
    n$\_$estimators & 20 \\
    subsample & 1 \\
    max$\_$depth & 3 \\
    objective & \textit{multi:softmax} \\
    eval$\_$metric & \textit{mlogloss} \\
\hline
\hline
\end{tabular}
\caption{XGBoost hyperparameters}
\label{Table1}
\end{table}

\begin{table}[h]
\centering
\textbf{CatBoost (CB)} \\
\begin{tabular}{l c }
\hline
\hline
Parameter & Value \\
\hline
    learning$\_$rate & 0.3 \\
    iterations & 30 \\
    depth & 5 \\
    loss$\_$function & \textit{MultiClass} \\
    eval$\_$metric & \textit{MultiClass} \\
\hline
\hline
\end{tabular}
\caption{CatBoost hyperparameters}
\label{Table2}
\end{table}

\begin{table}[h]
\centering
\textbf{Random Forest (RF)} \\
\begin{tabular}{l c }
\hline
\hline
Parameter & Value \\
\hline
    n$\_$estimators & 30 \\
    max-depth & 5 \\
\hline
\hline
\end{tabular}
\caption{Random Forest hyperparameters}
\label{Table3}
\end{table}

\begin{figure}[!ht]
    \centering
     \includegraphics[width = 0.8\linewidth]{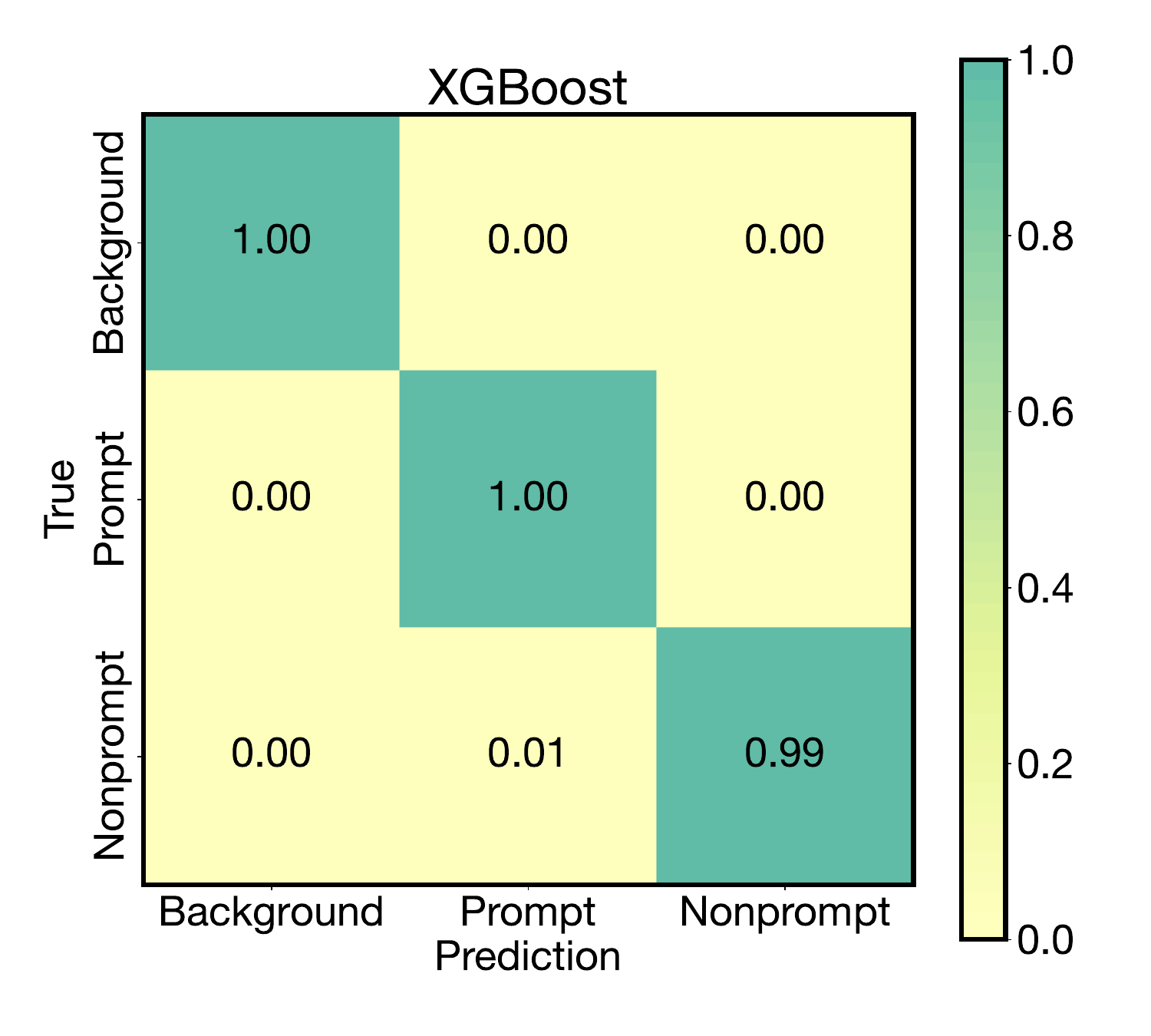}
    \includegraphics[width = 0.8\linewidth]{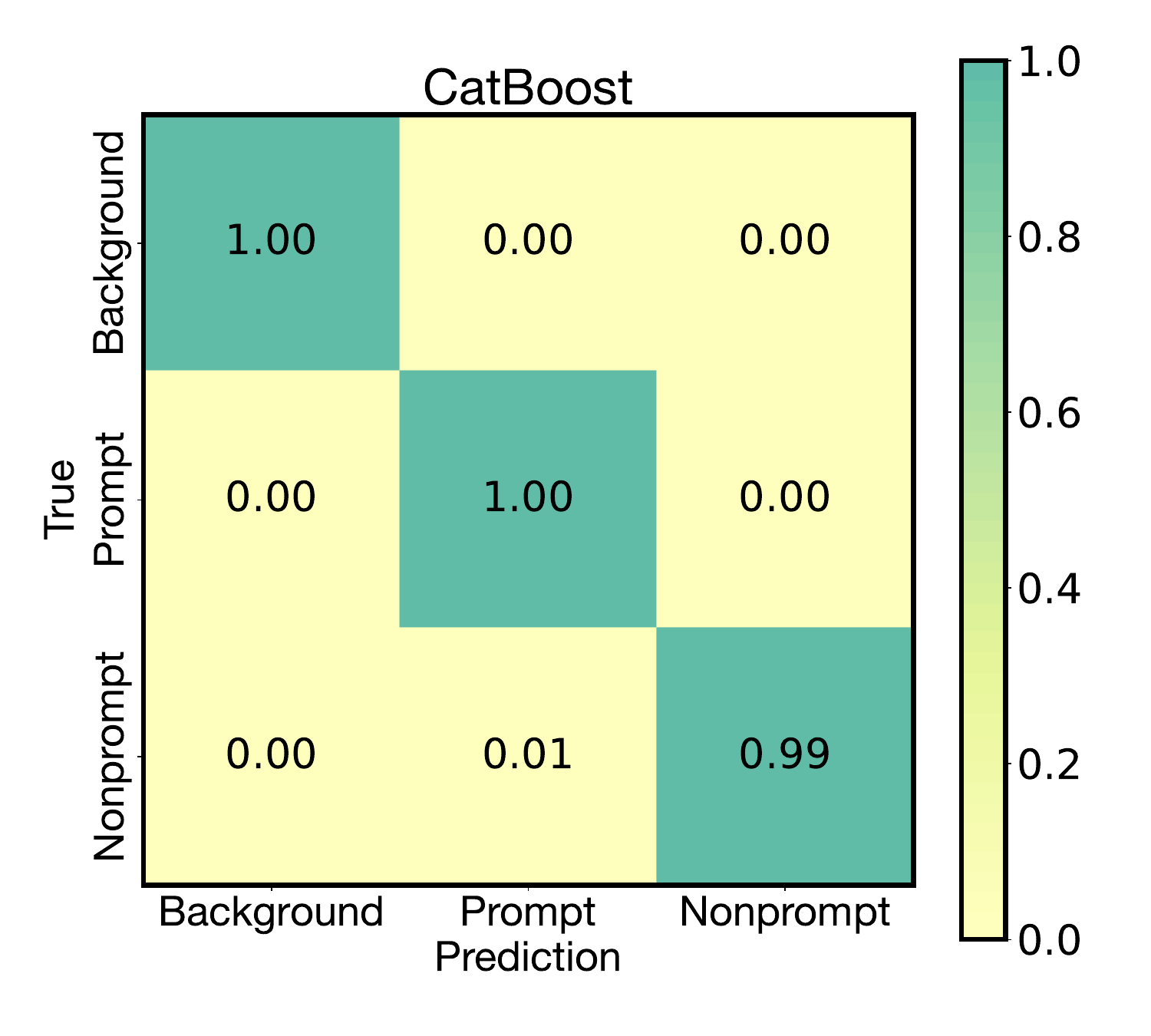}
    \includegraphics[width = 0.8\linewidth]{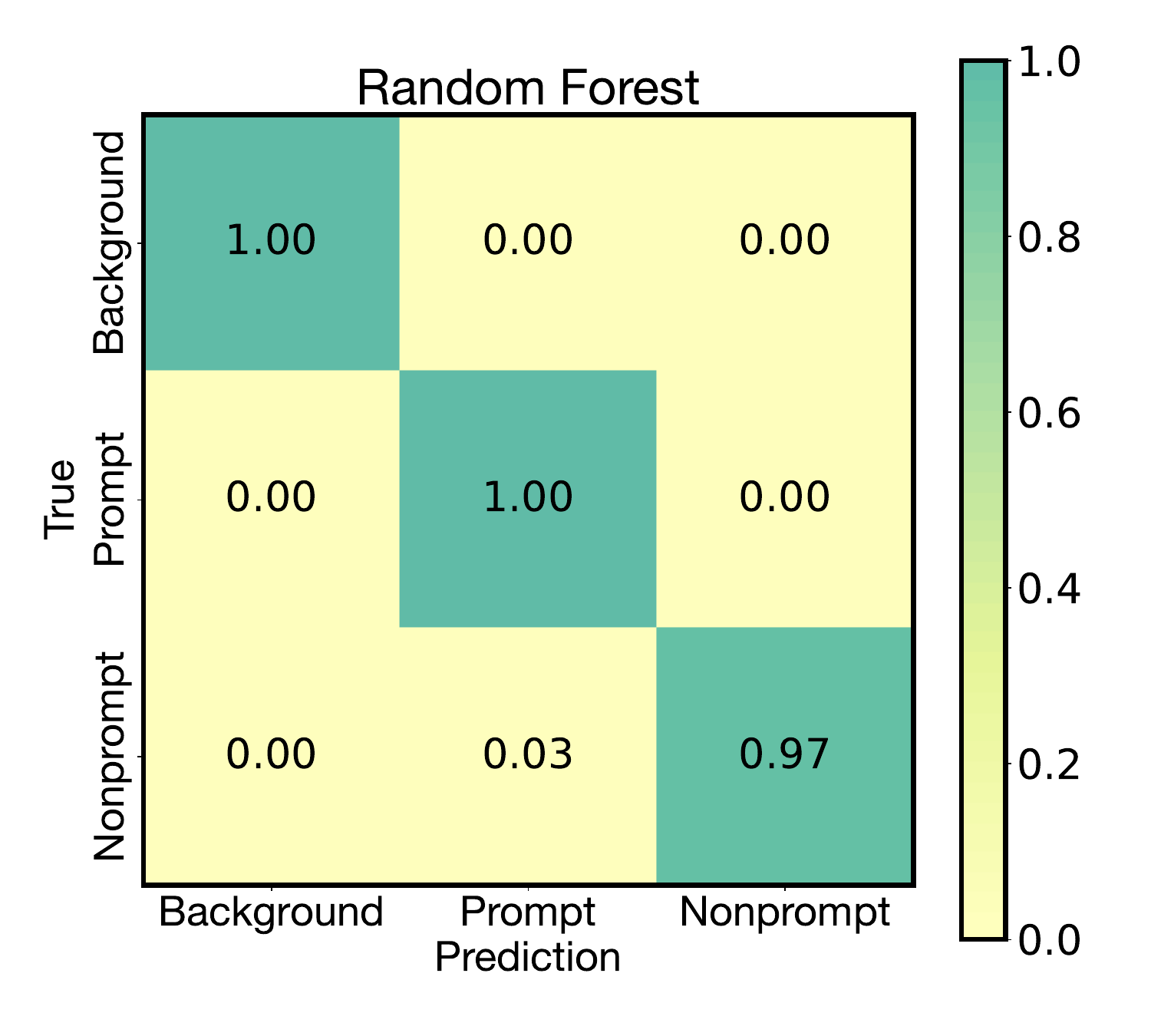}
    \caption{Confusion matrix for XGBoost (upper), CatBoost (middle), and Random Forest (lower), respectively. It represents the accuracy and discrepancy of the machine-learning models to predict the target classes.}
    \label{fig:CM}
\end{figure}

In XGBoost, the \textit{booster} decides the type of model that runs at each iteration. The \textit{gbtree} booster uses tree-based models. The \textit{learning rate} is a configurable hyperparameter that determines how much the weights in the model are adjusted during training. A higher \textit{learning rate} means the model learns faster, which could lead to overshooting the optimal solution. Conversely, a lower \textit{learning rate} means the model learns slower, which could lead to a more precise solution but at the cost of more Central Processing Unit (CPU) time. The \textit{n$\_$estimators}  parameter refers to the number of gradient-boosted trees that are used in the model. The \textit{subsample} parameter is used to control the fraction of the total training data that the model will use before it starts building trees. However, in our case the \textit{subsample} parameter is set to 1, allowing the model to use all the training data. The \textit{max$\_$depth} parameter decides the maximum depth of the tree. Increasing this parameter will make the model more complex and may lead to overfitting. The \textit{objective} parameter specifies the learning task and the corresponding learning objective. Setting the \textit{objective} as \textit{multi:softmax} tells the model that it is a multi-class classification problem. The softmax function is used to convert the output of the model into probability distributions over the classes. In XGBoost, the \textit{eval$\_$metric} parameter is utilized to define the evaluation metrics for the validation data. The choice of the evaluation metric heavily influences how the performance of a model is measured and compared. Here, \textit{mlogloss} refers to multi-class logarithmic loss, a loss function employed for multiclass classification problems. It is a negative logarithm of the predicted probability of the true class, the closer the probability is to 1, the smaller the output of the \textit{mlogloss}. Conversely, if the predicted probability of the true class is small (i.e., the prediction is likely to be incorrect), the \textit{mlogloss} value would be large.

In CatBoost, the first hyperparameter is the \textit{learning rate}, which we have kept at a value of 0.3. The second hyperparameter, \textit{iteration}, is used to control the number of trees to be built. Each iteration corresponds to a new tree being added to the model. Here, \textit{depth} corresponds to the maximum depth of the trees the algorithm is allowed to build. The \textit{loss$\_$functio}n and \textit{eval$\_$metric} are both taken as \textit{MultiClass}. This is the metric usually used for the training and evaluation of the model for a multi-class classification problem.

In a Random Forest model, the \textit{n$\_$estimators} parameter determines the count of trees in the forest. The model’s final prediction is derived by taking the average of the predictions from each tree. Although increasing the tree count can enhance the model’s effectiveness, it may also escalate the computational demand of the model. The \textit{max$\_$depth} serves the similar purpose of deciding the maximum depth of the trees in the model. All other hyperparameters that are not mentioned here are kept at their default values.

For our study, we start from the default values of the hyperparameters and progressively adjust them for a better result. In the case of all the ML algorithms used in this study, we can increase the \textit{n$\_$estimators} or \textit{max$\_$depth} to enhance the model performance, however, the prediction accuracy saturates with increasing the values of these
two hyperparameters leading to a considerable amount of CPU time. Thus, we strike a balance between the accuracy of the models and the computation time.

\subsection{Quality assurance}

\begin{figure}[!ht]
    \centering
    \includegraphics[width = 0.8\linewidth]{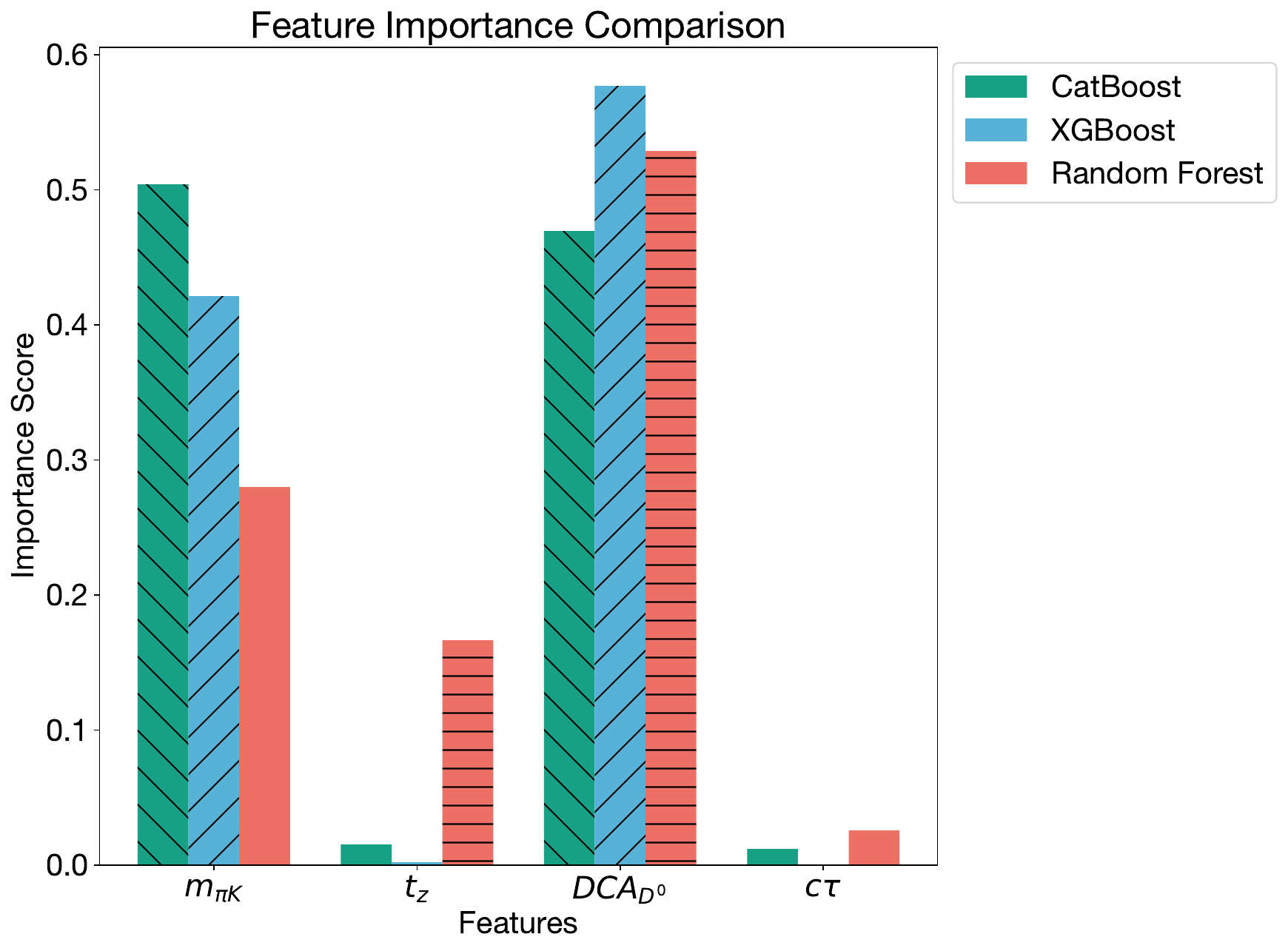}
    \caption{Comparison of Importance Score for the input variables, invariant mass ($m_{\pi K}$), pseudoproper time ($t_{z}$), pseudoproper decay length ($c\tau$), and distance of closest approach (\dca) for three different machine-learning algorithms.}
    \label{fig:IS_comp}
\end{figure}

After training the models, we proceed to evaluate them on a testing data set to check their classification accuracy. This tells us whether we can rely upon the trained models or not. 
For this classification problem, we use the confusion matrix to benchmark the ML models. In addition, a plot with the importance score of each input feature is shown for the three ML models, which depicts the relative importance of an input feature for the classification task. The relative importance of an input feature may vary from one model to the other.

\begin{figure*}[!ht]
    \centering
    \includegraphics[width = 0.8\linewidth]{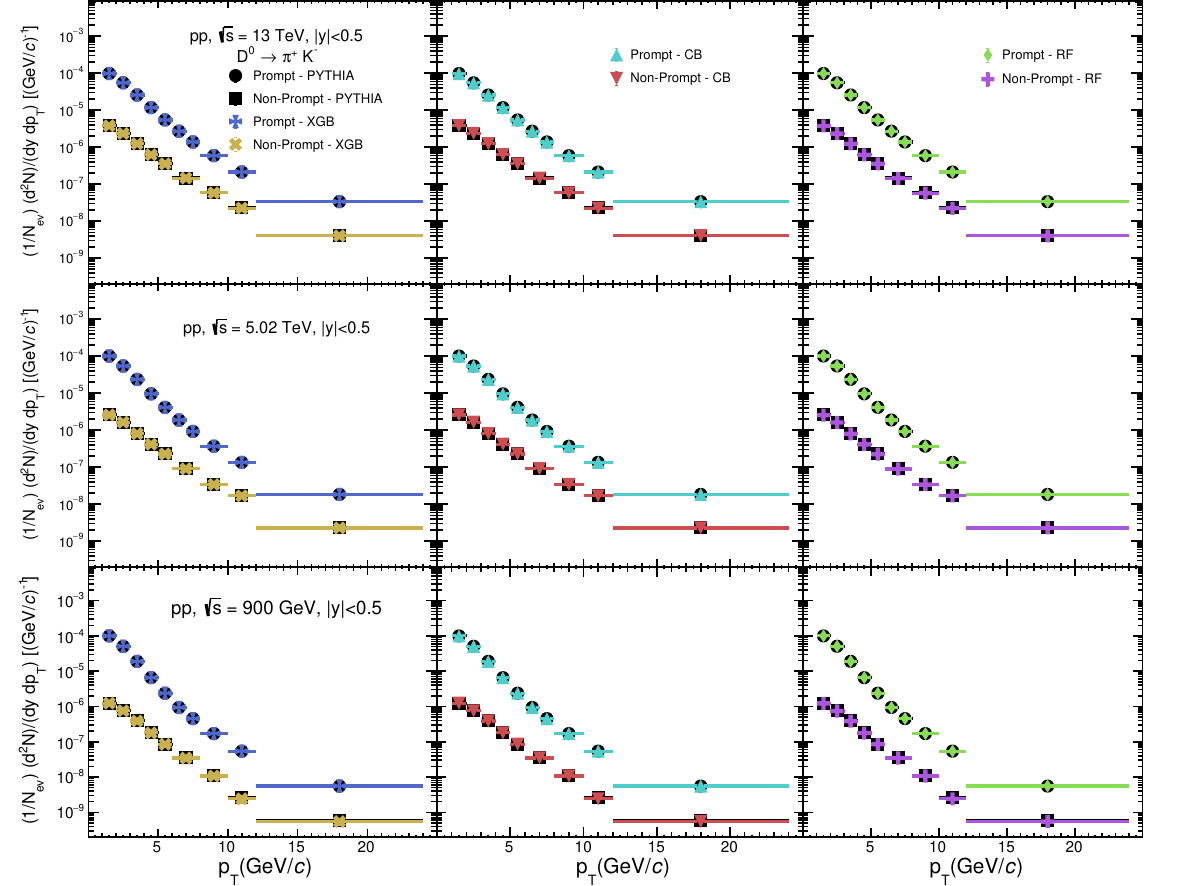}
    \caption{Transverse momentum spectra of prompt and nonprompt $D^{0}$ mesons at three different energies predicted by three different machine-learning algorithms. The first row depicts the normalized prompt and nonprompt $D^{0}$ yields in $pp$ collisions at $\sqrt{s} = 13~\rm{TeV}$. The second row presents the normalized yield at $\sqrt{s} = 5.02~\rm{TeV}$, and the third row illustrates the normalized yield at $\sqrt{s} = 900~\rm{GeV}$.}
    \label{fig:pT_spectra}
\end{figure*}

In Figure~\ref{fig:CM}, the confusion matrix for target classes, such as prompt, nonprompt, and background, for different ML algorithms used in this study is shown. The confusion matrix, or error matrix, is an essential benchmark in understanding the model performance. We plot the fraction of true pair counts from PYTHIA8 in the $Y$-axis and the fraction of predicted pair counts from ML models in the $X$-axis. The numbers shown inside the boxes represent the corresponding fraction of $\pi K$ pairs. All three ML models are found to separate the background pairs with an accuracy of 100\%. However, while separating prompt $D^{0}$ from the nonprompt ones, the XGBoost and CatBoost models have a better accuracy of 99\% as compared to the Random Forest model, which has an accuracy of 97\%. This means that the XGBoost and CatBoost models tag 1\% of the nonprompt $D^{0}$ mesons as prompt $D^{0}$ while this is 3\% in the Random Forest model. However, due to the imbalance between prompt and nonprompt classes, 1\% of nonprompt $D^{0}$ do not make a significant contribution to the prompt $D^{0}$ meson counts.
The magnitude of this misclassification is not prominent, and we expect to accurately extract other physics variables of the predicted prompt and nonprompt $D^0$ mesons, which are further discussed in Section~\ref{result_dis}.

\begin{figure*}[!ht]
    \centering
    \includegraphics[width = 0.8\linewidth]{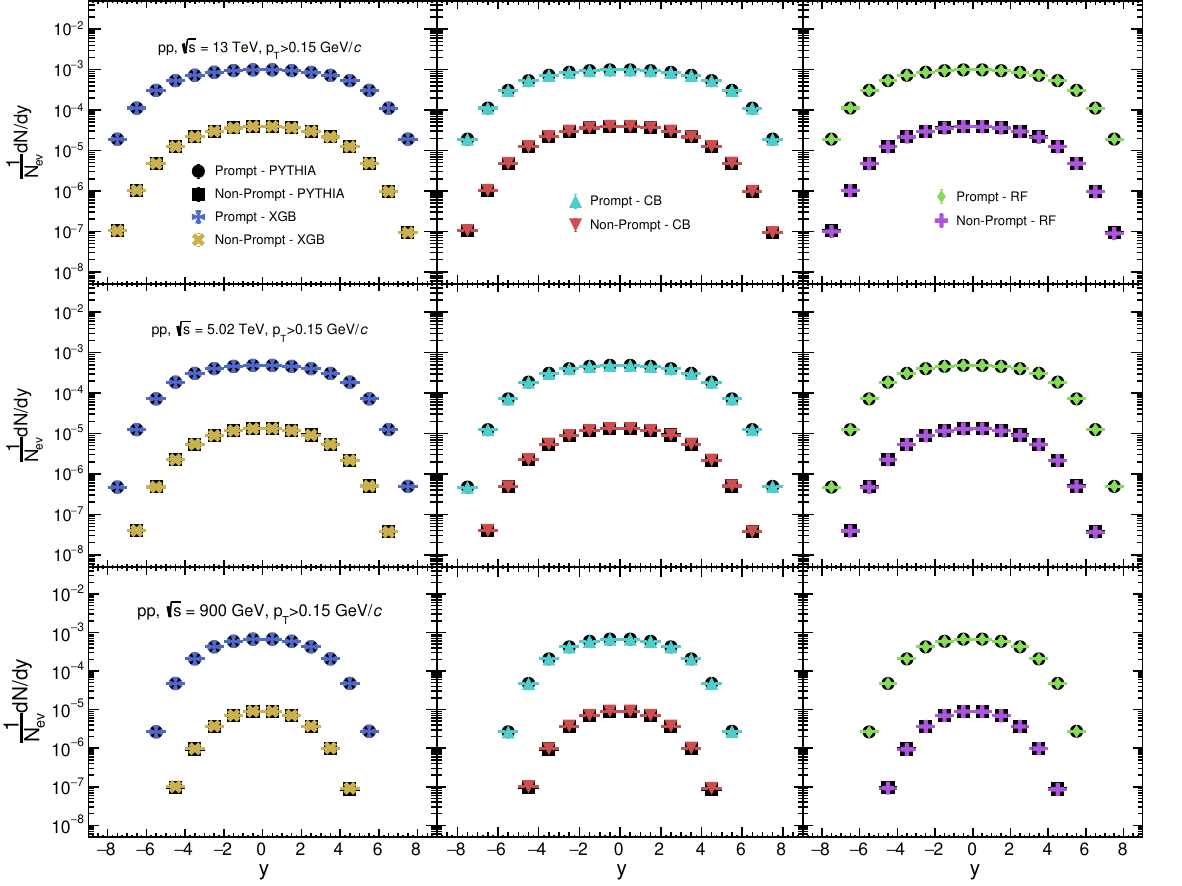}
    \caption{Rapidity spectra of prompt and nonprompt $D^{0}$ mesons at different energies predicted by three different machine-learning algorithms. The first row shows the normalized prompt and nonprompt $D^{0}$ yield in $pp$ collisions at center-of-mass energy, $\sqrt{s} = 13~\rm{TeV}$. The second row displays the normalized yield at $\sqrt{s} = 5.02~\rm{TeV}$, while the third row represents the normalized yield at $\sqrt{s} = 900~\rm{GeV}$.}
   \label{fig:y_spectra}
\end{figure*}

The importance score, or feature importance, is a score assigned to each input feature based on how useful it is in making a model prediction. It depends on the number of times the input feature is used in splitting a node. By looking at the importance score, one can figure out the most and least relevant features of the dataset for a particular ML model. In Figure~\ref{fig:IS_comp}, we show the importance score of the input features, $m_{\pi K}$, $t_{z}$, \dca, and $c\tau$. For all three ML models, the input features $m_{\pi K}$ and \dca possess the highest importance score. This signifies that these two input features carry the maximum information used in separating the background, prompt, and nonprompt classes. However, one can observe that the XGBoost model learns only from $m_{\pi K}$ and \dca explicitly. In contrast, the Random Forest model learns mostly from \dca, but also gives significant importance to $m_{\pi K}$ and $t_{z}$. The CatBoost model learns mostly from $m_{\pi K}$ and \dca; however, still uses input from $t_{z}$ and $c\tau$ for splitting the nodes.  

Beyond this point, for the rest of the text and figures, we use the following abbreviations: XGB for XGBoost, CB for CatBoost, and RF for Random Forest.

\section{Results and Discussion}
\label{result_dis}
\subsection{Transverse momentum and rapidity spectra}
Figure~\ref{fig:pT_spectra} shows the $p_{\rm{T}}$-differential yield of prompt and nonprompt $D^{0}$ meson in midrapidity, $|y|<0.5$, in $pp$ collisions at three different center-of-mass energies, i.e., $\sqrt{s}=13~\text{TeV}$ (upper), $\sqrt{s}=5.02~\text{TeV}$ (middle), and $\sqrt{s}=900~\text{GeV}$ (lower). We reconstruct $D^{0}$ meson through its hadronic decay channel, i.e., $D^{0}\rightarrow K^{-} \pi^{+}$. The plots include the predictions from XGB (left), CB (center), and RF (right). The PYTHIA8-generated spectra for the respective energies are also shown. All three ML models are trained with a minimum bias dataset of $pp$ collisions at $\sqrt{s} = 13~\rm{TeV}$ simulated with PYTHIA8 and then applied to $pp$ collisions at lower collision energies. One can observe that the yield of nonprompt $D^{0}$ is significantly less in the whole $p_{\rm T}$ region, owing to the smaller production probability of beauty hadrons due to their higher masses.
However, as one moves towards a higher $p_{\rm T}$ region, it can be seen that the $p_{\rm T}$-spectra curves from prompt and nonprompt $D^{0}$ mesons slightly approach each other. This indicates that the yield of nonprompt $D^{0}$ relative to the prompt $D^{0}$ meson increases with an increase in $p_{\rm T}$.
All three models are found to predict the normalized $D^{0}$ yield for energies $\sqrt{s} = 5.02~\rm{TeV}$ and $\sqrt{s} = 900~\rm{GeV}$ reasonably well. It is observed that the ML models are quite successful in predicting the $p_{\rm T}$-differential yield at different collision energies. Thus, they appear to retain the collision energy dependence. The ability of the models to learn and preserve the energy dependence of prompt and nonprompt $D^0$ production highlights their robustness and accuracy. This is primarily due to their learning that is largely influenced by two factors, the invariant mass ($m_{\pi K}$) and the distance of the closest approach (\dca), which are independent of $\sqrt{s}$.

Figure~\ref{fig:y_spectra} shows the rapidity spectra of prompt and nonprompt $D^{0}$ reconstructed from candidates with $p_{\rm T}>0.15$ GeV/c in minimum bias $pp$ collisions at $\sqrt{s} = 13~\rm{TeV}$ (upper), $\sqrt{s} = 5.02~\rm{TeV}$ (middle), and $\sqrt{s} = 900~\rm{GeV}$ (lower). The results from PYTHIA8, XGB, CB, and RF are shown. The energy dependence of the width of the rapidity spectra is noticeable, and the differences can be clearly observed by comparing the highest and lowest center-of-mass energies. In addition, the width of the rapidity spectra of the prompt $D^0$ meson is always greater than that of the nonprompt case at any given energy. For $\sqrt{s} = 13~\rm{TeV}$, the midrapidity region for the prompt $D^{0}$ seems flat in log-scale in the range, $|y|\lesssim 3$; however, this flat region for the prompt $D^{0}$ decreases with decreasing the collision energy. For $\sqrt{s} = 5.02~\rm{TeV}$, the flat region confides in a slightly smaller rapidity range of $|y| \lesssim 2$. This region shrinks even more for $\sqrt{s} = 900~\rm{GeV}$ where a smaller plateau exists only within $|y| \lesssim 1$. Moreover, this flat midrapidity plateau is much smaller for the nonprompt $D^{0}$ meson, as evident from the plots. The flat region is almost non-existent for $\sqrt{s} = 900~\rm{GeV}$. However, it extends to a range of $|y| \lesssim 2$ for $\sqrt{s} = 13~\rm{TeV}$. From Figs.~\ref{fig:pT_spectra} and \ref{fig:y_spectra}, we notice that all three ML models predict a similar level of accuracy in the yield of $D^0$ meson as a function of transverse momentum and rapidity. Consequently, beyond this point in the text, for the sake of clarity in the plots, we will be only focusing on the predictions from the XGB model and comparing them with PYTHIA8 results. The XGB model shows a higher degree of accuracy compared to the RF model in this scenario. Even though CB shows a similar degree of accuracy, adding predictions from two machine-learning algorithms only clutters the figures.

\begin{figure}
    \centering
    \includegraphics[width = 0.8\linewidth]{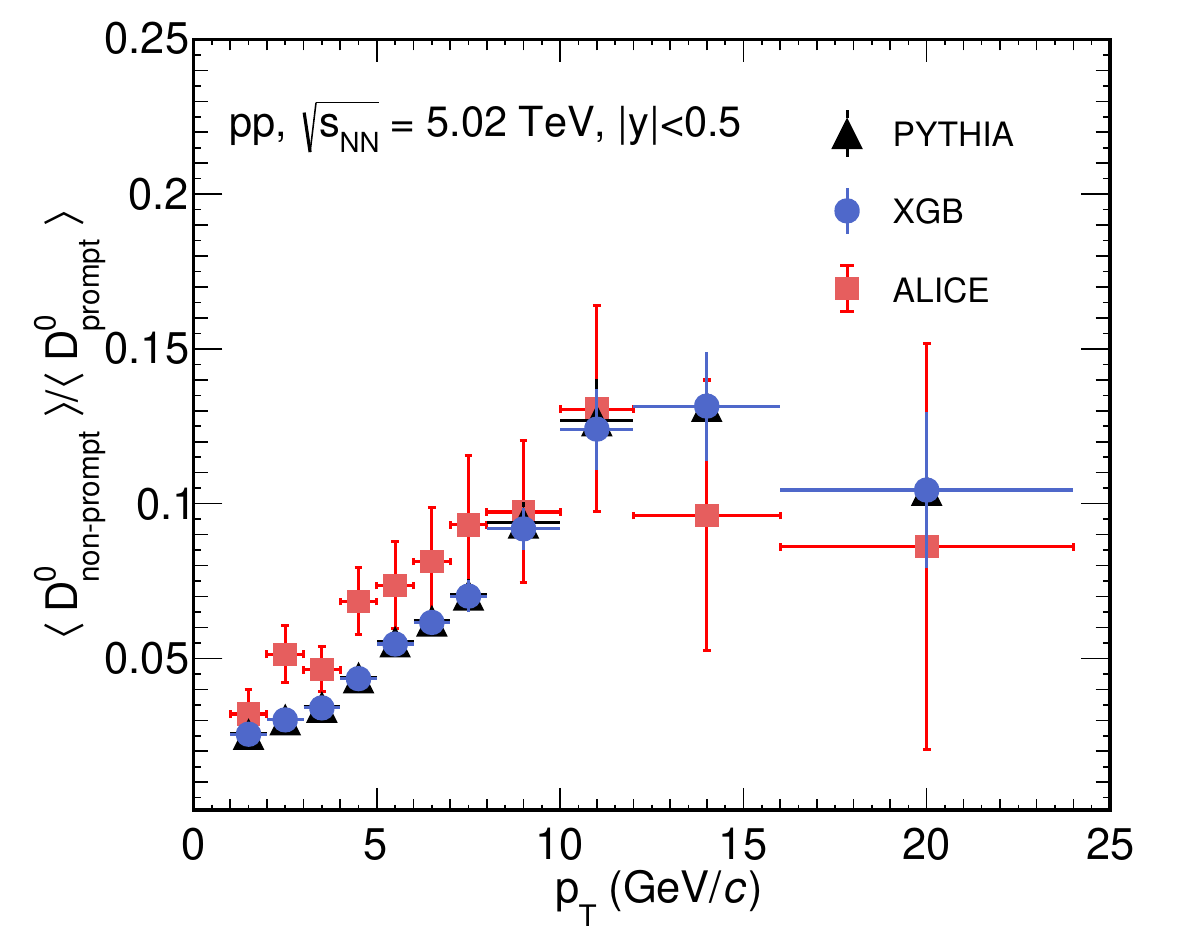}
    \caption{Nonprompt to prompt $D^0$ meson ratio in minimum bias $pp$ collisions at $\sqrt{s} = 5.02~\rm{TeV}$ from PYTHIA8 compared with ALICE results and predictions from XGB~\cite{ALICE:2021mgk}.}
    \label{fig:non-prompt_to_prompt_ALICE}
\end{figure}

\begin{figure}
    \centering
    \includegraphics[width = 0.8\linewidth]{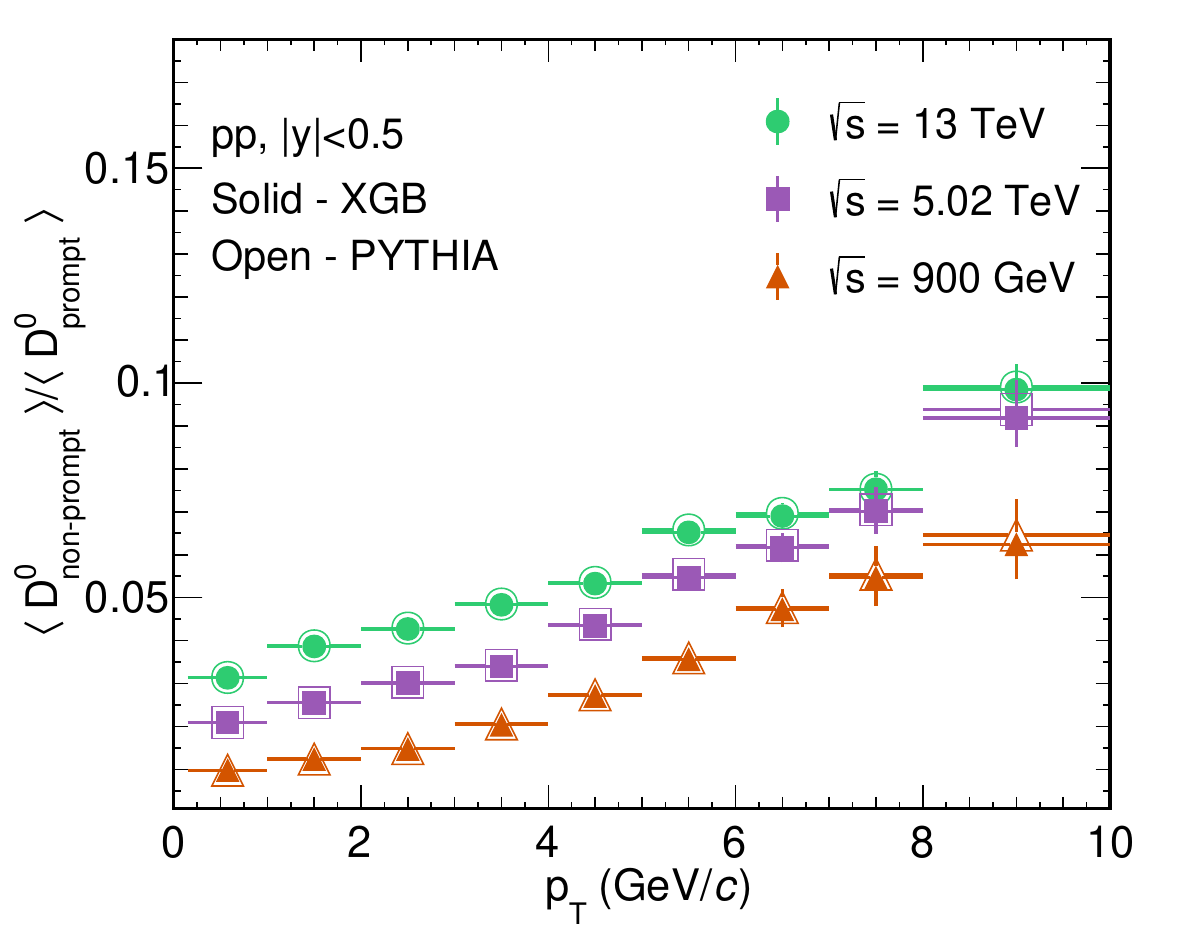}
    \caption{Nonprompt to prompt $D^0$ meson ratio for three different center-of-mass energies from PYTHIA8 compared with the predictions from XGB in minimum bias $pp$ collisions.}
    \label{fig:non-prompt_to_prompt}
\end{figure}

\subsection{Nonprompt to prompt ratio and self-normalized yield of $D^{0}$ meson}
Figure~\ref{fig:non-prompt_to_prompt_ALICE} presents the ratio of nonprompt to prompt $D^{0}$ yield, at midrapidity, $|y|<0.5$, in minimum bias $pp$ collisions at $\sqrt{s} = 5.02~\rm{TeV}$, as a function of $p_{\rm T}$. This ratio essentially tells us about the relative yield of $D^0$ mesons coming from beauty hadron decays, compared to the direct charm hadron production. We compare the XGB predictions with the ALICE~\cite{ALICE:2021mgk} results. From the plot, it is observed that PYTHIA8 underestimates the experimental results at lower-$p_{\rm T}$ and starts to approach the experimental results only towards the higher-$p_{\rm T}$ bins. However, the overall trend of PYTHIA8 is similar to that of the experimental findings. Again, the nonprompt to prompt $D^0$ yield ratio increases linearly up to $p_{\rm{T}} = 12~\rm{GeV}/\textit{c}$. This indicates that the probability of charm hadron production from beauty decays increases linearly with $p_{\rm{T}}$. The linear trend holds good up to a certain $p_{\rm{T}}$ range. Similar results are also reported for charmonium states \cite{ALICE:2021edd, Prasad:2023zdd}. However, the increase in nonprompt charmonium states as a function of $p_{\rm T}$ is much higher than that of the open-charm states ~\cite{Prasad:2023zdd}. Moreover, for $p_{\rm{T}}>12~\rm{GeV}/\textit{c}$, the ALICE data is uncertain with larger error bars, and the trend appears to become independent of $p_{\rm{T}}$. The predictions from XGB are found to be in line both qualitatively and quantitatively with the PYHTIA8 true values.

Figure~\ref{fig:non-prompt_to_prompt} shows the nonprompt to prompt $D^{0}$ ratio in minimum bias $pp$ collisions at three different center-of-mass energies, i.e., $\sqrt{s} = 13~\rm{TeV}$, $5.02~\rm{TeV}$, and  $900~\rm{GeV}$. One can clearly notice the increase in the ratio with increasing $p_{\rm T}$ across all the collision energies. However, we observe an energy-dependent hierarchy in the ratio, as the charm production from beauty decays compared to the direct charm production is minimum for $\sqrt{s} = 900~\rm{GeV}$ and maximum for the case of $\sqrt{s} = 13~\rm{TeV}$. In addition, towards higher $p_{\rm{T}}$, we see the rise of the ratio, indicating an increase in the beauty hadron production leading to an enhancement of the nonprompt yield.

\begin{figure}
    \centering
    \includegraphics[width = 0.8\linewidth]{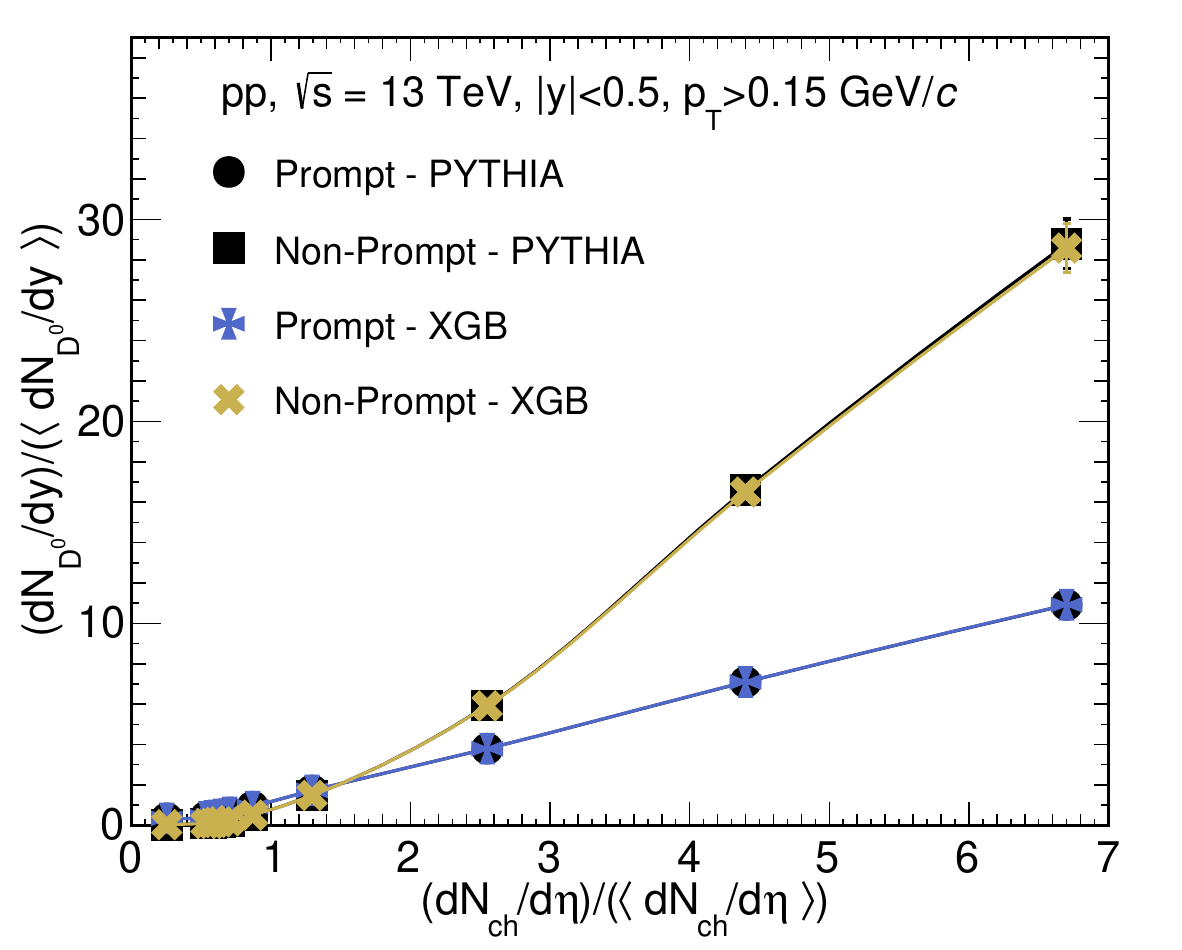}
    \includegraphics[width = 0.8\linewidth]{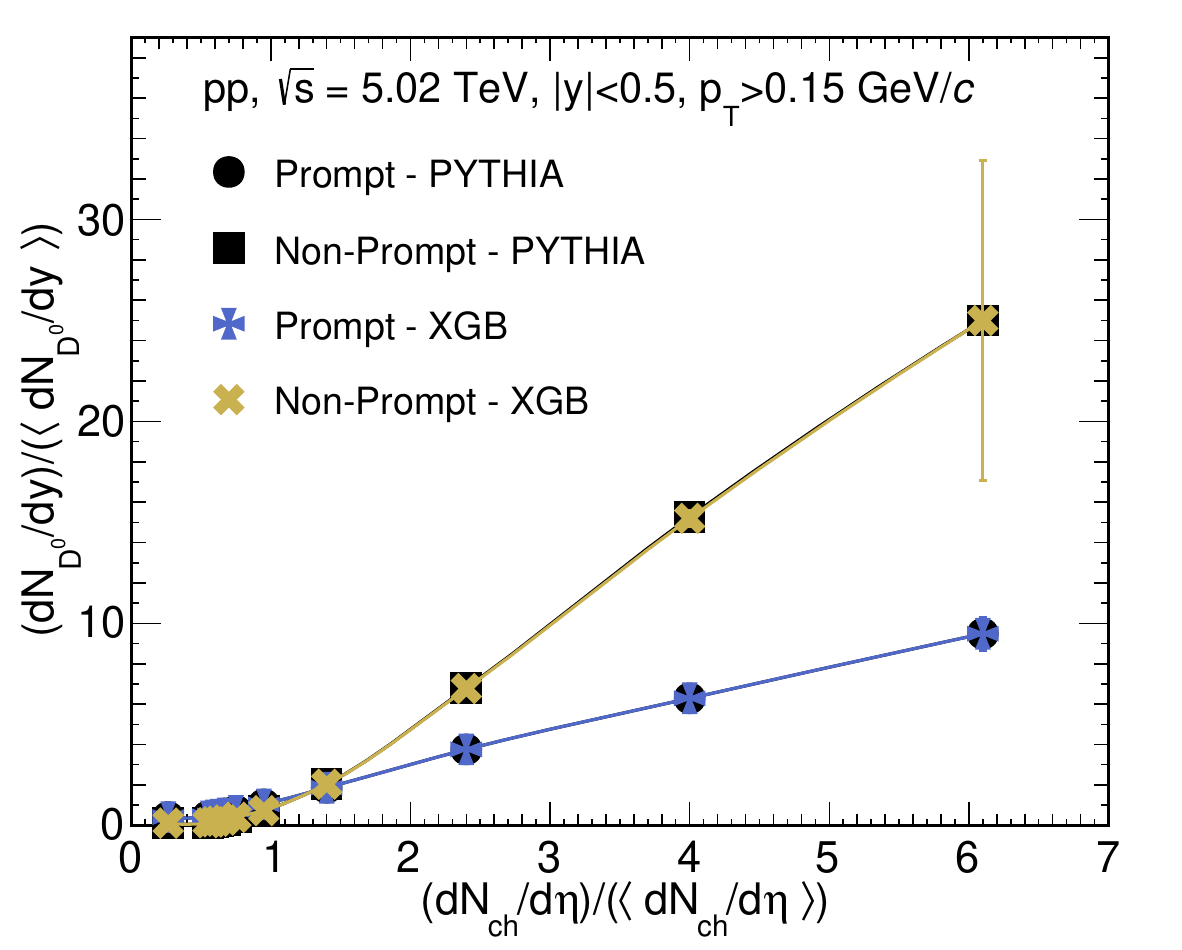}
    \includegraphics[width = 0.8\linewidth]{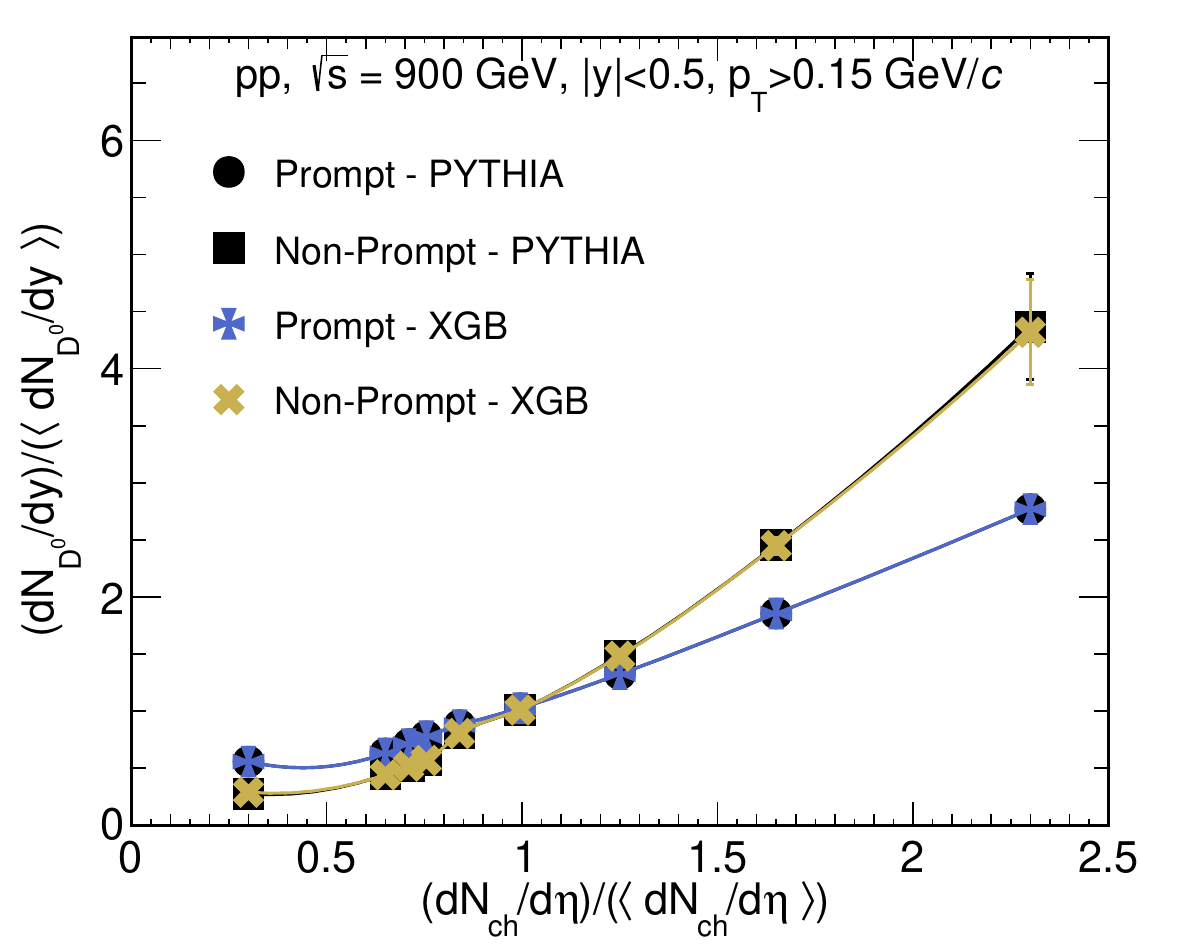}
    \caption{Self-normalized $p_{\rm{T}}$ integrated prompt and nonprompt $D^{0}$ meson yield at midrapidity ($|y|<0.5$) as a function of normalized charged-particle multiplicity in minimum bias $pp$ collisions at $\sqrt{s} = 13~\rm{TeV}$ (upper), $\sqrt{s} = 5.02~\rm{TeV}$ (middle), and $\sqrt{s} = 900~\rm{GeV}$ (lower). The charged-particle multiplicity is obtained within the ALICE-V0 detector acceptance.}
    \label{fig: self-normalized yield}
\end{figure}

Figure~\ref{fig: self-normalized yield} shows the self-normalized $p_{\rm{T}}$ integrated yield of prompt and nonprompt $D^{0}$ meson in midrapidity ($|y|<0.5$) as a function of normalized charged-particle multiplicity in minimum bias $pp$ collisions at $\sqrt{s} = 13~\rm{TeV}$ (upper), $\sqrt{s} = 5.02~\rm{TeV}$ (middle), and $\sqrt{s} = 900~\rm{GeV}$ (lower). The charged-particle multiplicity is obtained within the ALICE-V0 detector acceptance which covers the intervals $2.8<\eta<5.1$ (V0A) and $-3.7<\eta<-1.7$ (V0C). The charged-particle multiplicity used for the normalized yield selection is the coincidence signal of V0A and V0C. The selection of $D^0$ meson and charged particle multiplicity in two different rapidity regions is to reduce the autocorrelation bias. The results include PYTHIA8 values and the prediction from the XGB model. We observe an almost linear rise for the prompt $D^{0}$ meson with respect to the charged particle multiplicity for all three collision energies. However, the self-normalized yield of nonprompt $D^{0}$ is significantly enhanced towards higher collision energy and follows a faster-than-linear trend with increasing charged-particle multiplicity. A similar trend for charmonium states (i.e. $J/\psi$) has been reported in the literature using PYTHIA8~\cite{Prasad:2023zdd}. For the plots shown here, XGB predictions closely follow the PYTHIA8 curves.

\begin{figure}
    \centering
    \includegraphics[width = 0.8\linewidth]{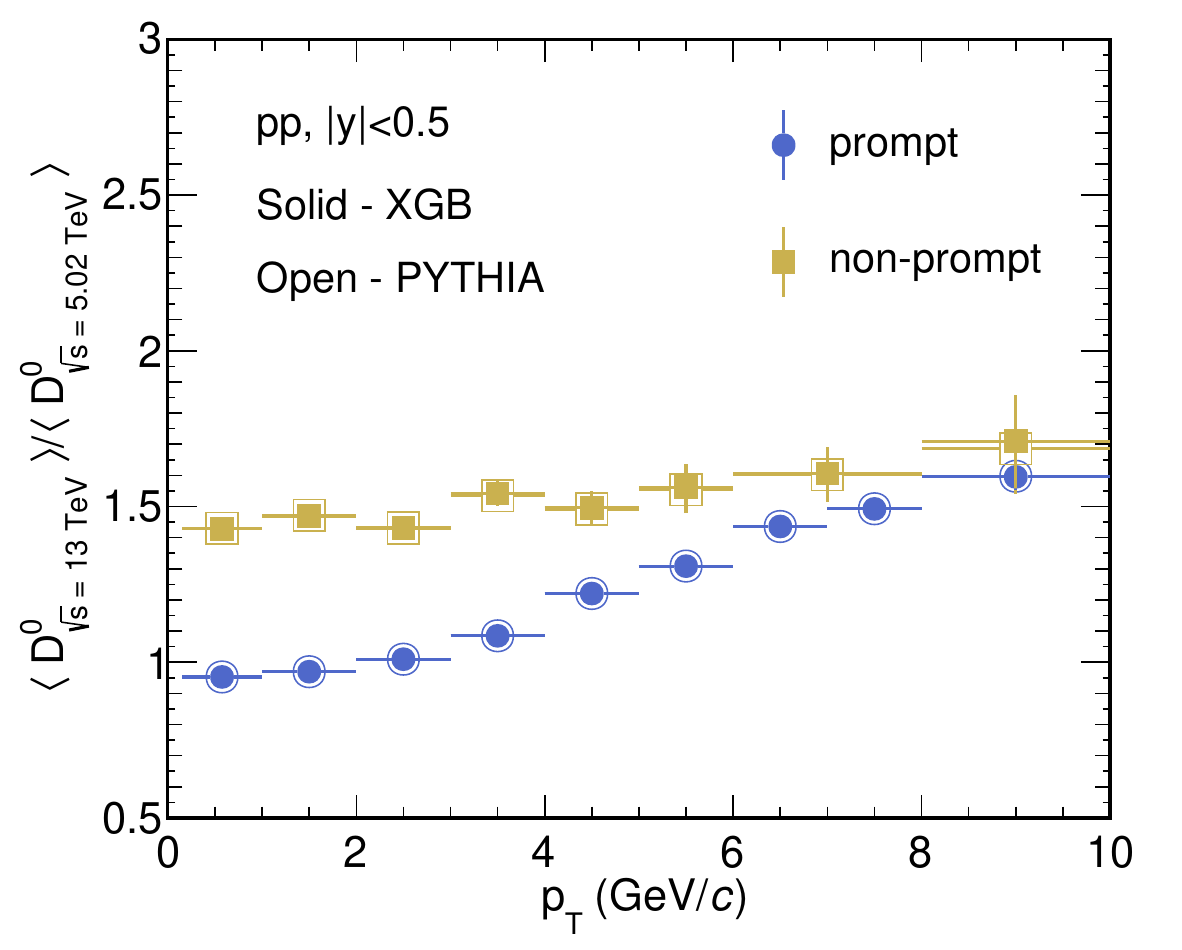}
    \includegraphics[width = 0.8\linewidth]{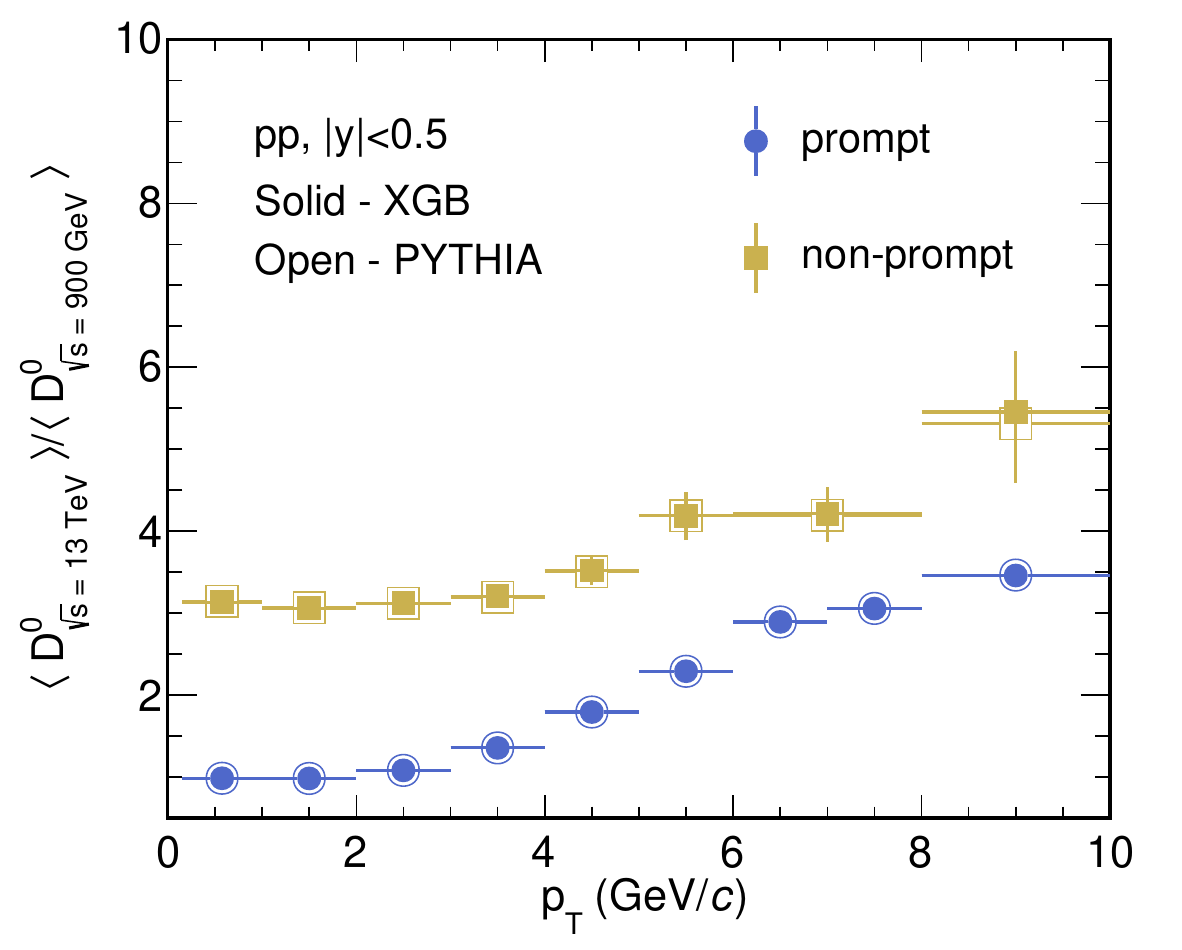}
    \caption{Ratio of $D^{0}$ yield in $pp$ collisions at $\sqrt{s} = 13$ and 5.02 TeV  (upper) and at $\sqrt{s} = 13~\rm{TeV}$ and $\sqrt{s} = 900~\rm{GeV}$ (lower) as a function of $p_{\rm{T}}$.}
    \label{fig:nrg_frac}
\end{figure}

Finally, we study the role of center-of-mass energy in $D^{0}$ meson production. We estimate the ratio of $D^{0}$ yield in two different energies. In the upper panel of Fig. \ref{fig:nrg_frac}, we plot the ratio of $D^{0}$ yield in $\sqrt{s} = 13~\rm{TeV}$ to $\sqrt{s} = 5.02~\rm{TeV}$. Here, for the prompt case, we notice a clear increase in the ratio with an increasing $p_{\rm{T}}$. However, we observe a flat trend throughout the whole $p_{\rm{T}}$ range for the nonprompt case. 
A similar trend has been observed recently at ALICE~\cite{ALICE:2024xln}. In addition, a higher value of the nonprompt than prompt ratio shows the abundant production of beauty hadrons at higher center-of-mass energy. In the lower panel, we plot the same ratio between $\sqrt{s} = 13~\rm{TeV}$ and $\sqrt{s} = 900~\rm{GeV}$. Because of a significant difference in collision energy, we observe the increasing trend of the ratio as a function of transverse momentum for both prompt and nonprompt cases. Furthermore, due to the higher difference in the center-of-mass energy, the absolute values of the ratio go up (lower panel compared to the upper panel of Fig. \ref{fig:nrg_frac}). Interestingly, XGB can predict the PYTHIA8 trends with very high accuracy.

In this study, we train the model with PYTHIA simulated $pp$ collisions, and the models can predict the results at different energies reasonably well. However, this would not be the case for $p$--Pb and Pb--Pb collisions as the final state charged particle multiplicity is comparatively much higher than the $pp$ collisions. Moreover, this steep increase in the multiplicity will affect particle production, which in turn changes the prompt and nonprompt production dynamics. 

\begin{figure}
    \centering
    \includegraphics[width = 0.8\linewidth]{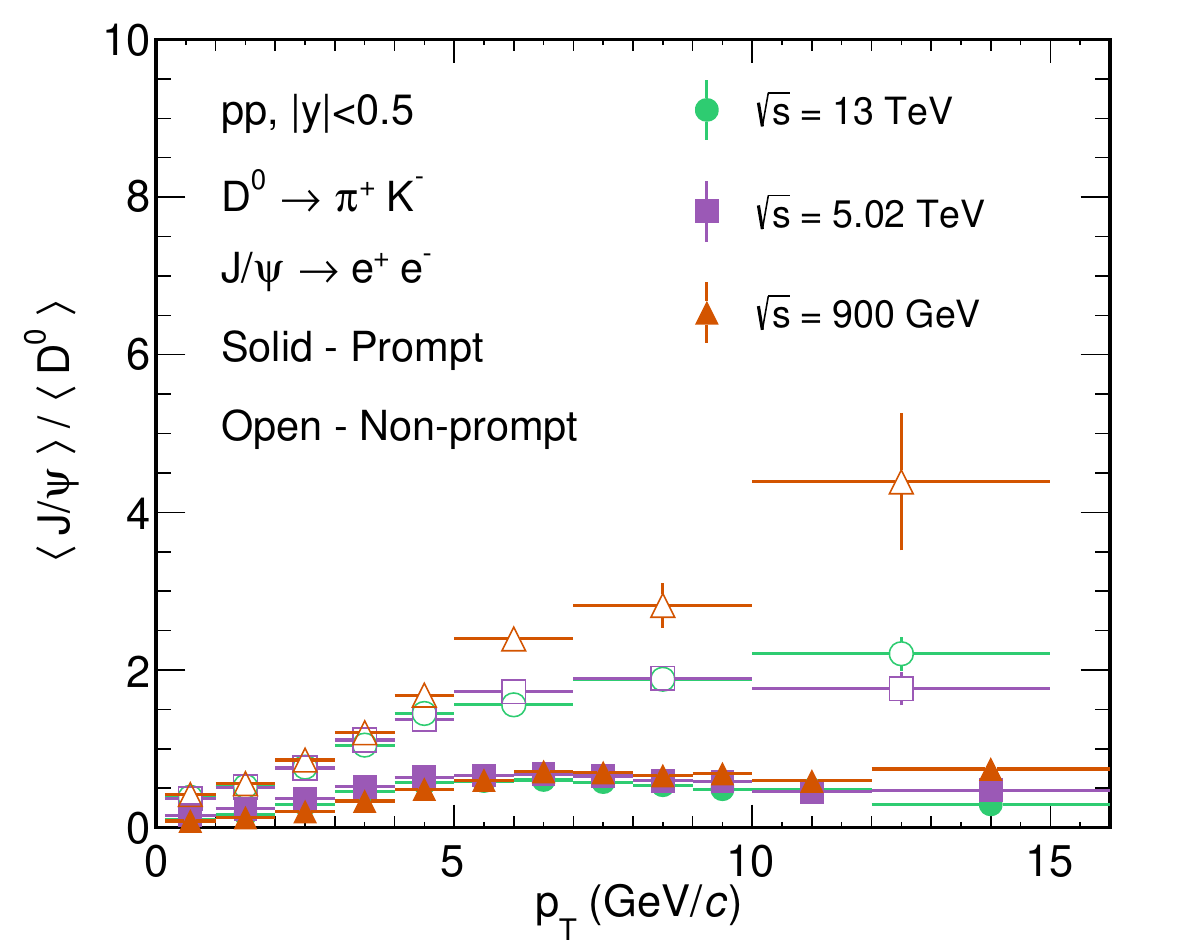}
    \includegraphics[width = 0.8\linewidth]{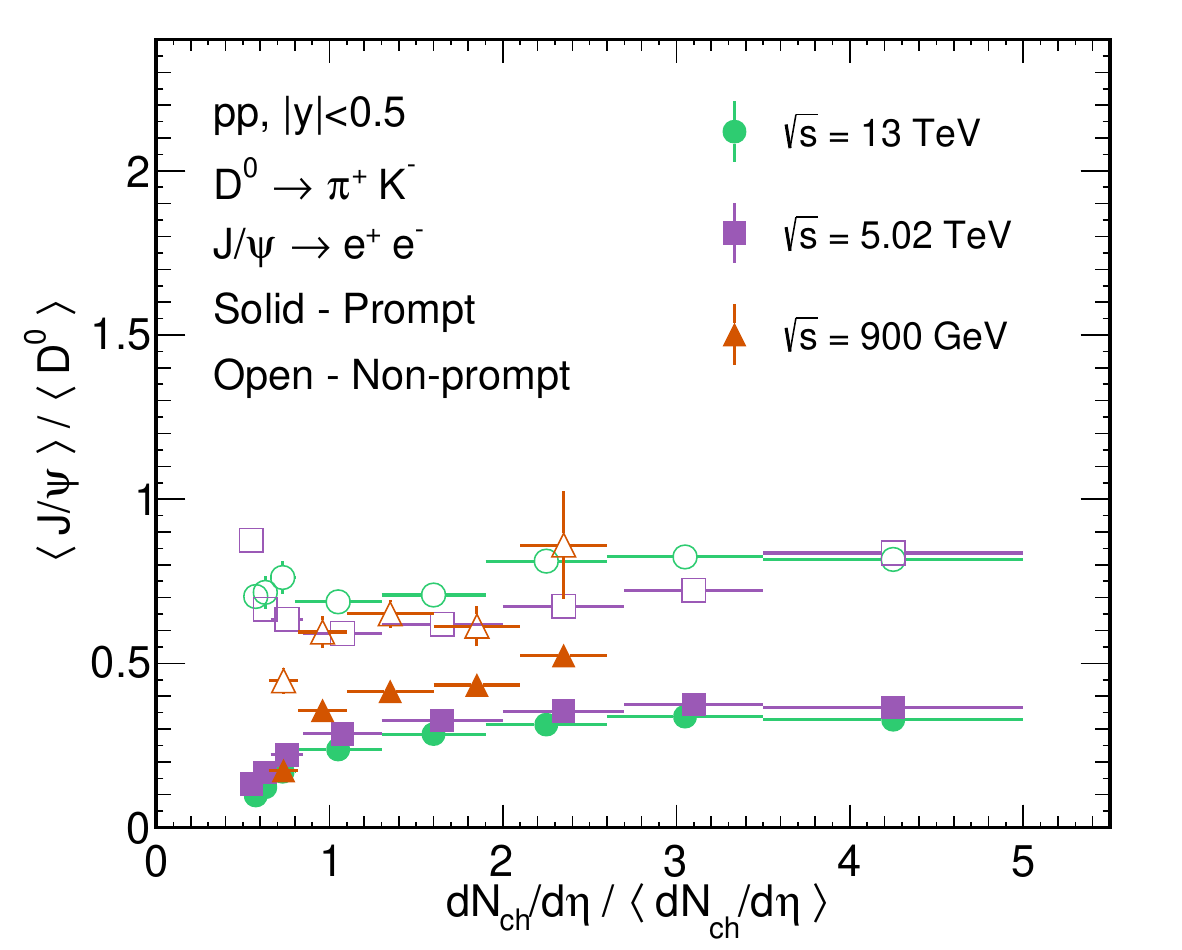}
    \caption{Upper: Normalized $J/\psi$ yield to $D^{0}$ yield ratio as a function of $p_{\rm T}$ in $pp$ collisions at $\sqrt{s} = 13$ and 5.02 TeV and $\sqrt{s} = 900~\rm{GeV}$. Lower: $p_{\rm{T}}$ integrated normalized $J/\psi$ yield to $D^{0}$ yield ratio as a function of normalized charged-particle multiplicity estimated within ALICE-V0 acceptance.}
    \label{fig: Jpsi_D0_ratio}
\end{figure}

\subsection{Ratio of charmonium to open-charm state}

It is interesting to study the production dynamics of charmonium states relative to open-charm states. In Fig.~\ref{fig: Jpsi_D0_ratio}, on the upper panel, the normalized $J/\psi$ to $D^{0}$ yield as a function of $p_{\rm T}$ is shown in minimum bias $pp$ collisions. To understand the contribution coming from the charm and beauty sectors, we estimate the ratio of prompt $J/\psi$ to prompt $D^{0}$, as well as the ratio of nonprompt $J/\psi$ to nonprompt $D^{0}$. We observe similar trends for the prompt and nonprompt cases up to $p_{\rm{T}}\simeq 5~\rm{GeV}/\textit{c}$, a rise in the ratio can be seen. However, the prompt $J/\psi$ to prompt $D^{0}$ ratio decreases slightly after $p_{\rm{T}}\simeq 5~\rm{GeV}/\textit{c}$ as compared to the nonprompt case, which remains flat. This trend is universal for all center-of-mass energies. It indicates that the relative number of $J/\psi$ increases as compared to $D^{0}$, with an increase in $p_{\rm T}$. One can notice that the ratio of nonprompt $J/\psi$ to nonprompt $D^{0}$ is higher than one, indicating a higher number of nonprompt $J/\psi$ compared to nonprompt $D^{0}$. 
Assuming that the same beauty hadrons contribute to the production of nonprompt $J/\psi$ and nonprompt $D^0$ mesons, $\langle J/\psi \rangle/\langle D^{0}\rangle>1$ for nonprompt case indicates that a beauty hadron would more likely to decay into a $J/\psi$ than to a $D^{0}$ mesons. In other words, the branching fraction of beauty hadrons decaying into $J/\psi$ is higher than their decay to $D^{0}$.
However, as expected, the ratio of prompt $J/\psi$ to prompt $D^{0}$ is less than one, owing to the larger mass of $J/\psi$.
In the lower panel, we present the $\langle J/\psi \rangle/\langle D^{0} \rangle$ as a function of normalized charged-particle multiplicity. Here, we observe two different trends for the prompt and nonprompt cases. We notice the nonprompt $J/\psi$ to $D^{0}$ ratio remains almost independent of normalized charged-particle multiplicity. However, for the prompt case, we notice a slight increase and then a flat trend in the ratio with the increase in the normalized charged-particle multiplicity. Additionally, there is a noticeable ordering in the prompt $J/\psi$ to $D^{0}$ ratio, where the ratio increases with a decrease in collision energy. This trend is consistent throughout all the charged-particle multiplicities.

\section{Summary}
\label{summ}
In this paper, we present a novel method for track-level (unbinned) identification and segregation of the prompt and nonprompt $D^{0}$ from the background pion-kaon pairs using machine learning algorithms. We use experimentally measurable topological variables as inputs, which include the invariant mass ($m_{\pi K}$), pseudoproper time ($t_{z}$), pseudoproper decay length ($c\tau$), and distance of closest approach (\dca). We train the XGB, CB, and RF models with data generated using PYTHIA8 for $pp$ collisions at $\sqrt{s} = 13~\rm{TeV}$. The XGB and CB models show an accuracy up to 99\% in separating prompt and nonprompt $D^{0}$ mesons; however, the RF model shows an accuracy of 97\%. The models are efficient and robust enough to predict the results even at lower collision energies: $\sqrt{s} = 5.02~\rm{TeV}$ and $\sqrt{s} = 900~\rm{GeV}$ in the complete transverse momentum and pseudorapidity region.

Also, to understand the production of prompt and nonprompt $D^{0}$ meson, we study the nonprompt to prompt ratio of $D^{0}$ yield as a function of transverse momenta. Furthermore, we study the self-normalized yield of $D^{0}$ meson, where we observe a nonlinear rising trend for the nonprompt $D^{0}$ as a function of normalized charged particle multiplicity.
In addition, we have incorporated predictions and results from several collision energies, which not only serve as a benchmark for the predictions from the machine learning models but also provide a collision energy dependence study of prompt and nonprompt $D^0$ mesons.
Finally, we explore the relative production of charmonium, $J/\psi$ to open-charm, $D^{0}$ states as a function of transverse momenta and charged-particle multiplicity. In all these studies, the predictions from XGB match the PYTHIA8 values quite well. This method has an advantage over the conventional methods as it can perform unbinned measurements for both prompt and nonprompt $D^{0}$ by directly tagging the decay daughters.

The ongoing ALICE Run 3 data taking with high luminosity and better detection capabilities would pave the way for several precise measurements for the charm and beauty sector. The separation of charm hadron topological production into prompt and nonprompt ones allows us to explore the beauty sector. The ability to separate the contribution from the beauty sector gives us a better understanding of the dynamics of the charmed hadron production, their interaction with the QGP, and the properties of the QCD medium. The use of machine learning algorithms can help us replace the traditional fitting procedures with improved track-level identification of the prompt and nonprompt production of charm hadrons. The production dynamics of prompt vs non-prompt
charmonium and open charm at the LHC energies using the ALICE upgrade would provide a test bench for QCD and
the study of multihadron production dynamics extending to the beauty sector at the subatomic level.

This study demonstrates the efficiency of
using machine learning techniques in the topological separation of open charm mesons using standalone PYTHIA8 Monte Carlo event simulation. In experiments, heavy-flavor measurements are greatly influenced by several track selection criteria for their decay candidates. This makes particle identification and reconstruction more challenging in real-life scenarios. Thus, to conclude the working and stability of the proposed models in this study, one should fully reconstruct the simulated events using an identical experimental detector setup, which can be performed using the GEANT3 transport package~\cite{Brun:1994aa}. This latter part of the analysis can be taken as an outlook of the present study.

\section{Acknowledgements}
K. G. acknowledges the financial support from the Prime Minister’s Research Fellowship (PMRF), Government of India. S.P. acknowledges the doctoral fellowships from the University Grants Commission (UGC), Government of India. The authors acknowledge the DAE-DST, Government of India funding under the Mega-Science Project—“Indian participation in the ALICE experiment at CERN” bearing Project No. SR/MF/PS-02/2021-IITI (E-37123). The authors sincerely acknowledge the usage of resources of the LHC grid Tier-3 computing facility at IIT Indore.

\section*{Appendix}
\begin{figure}
    \centering
    \includegraphics[width = 0.8\linewidth]{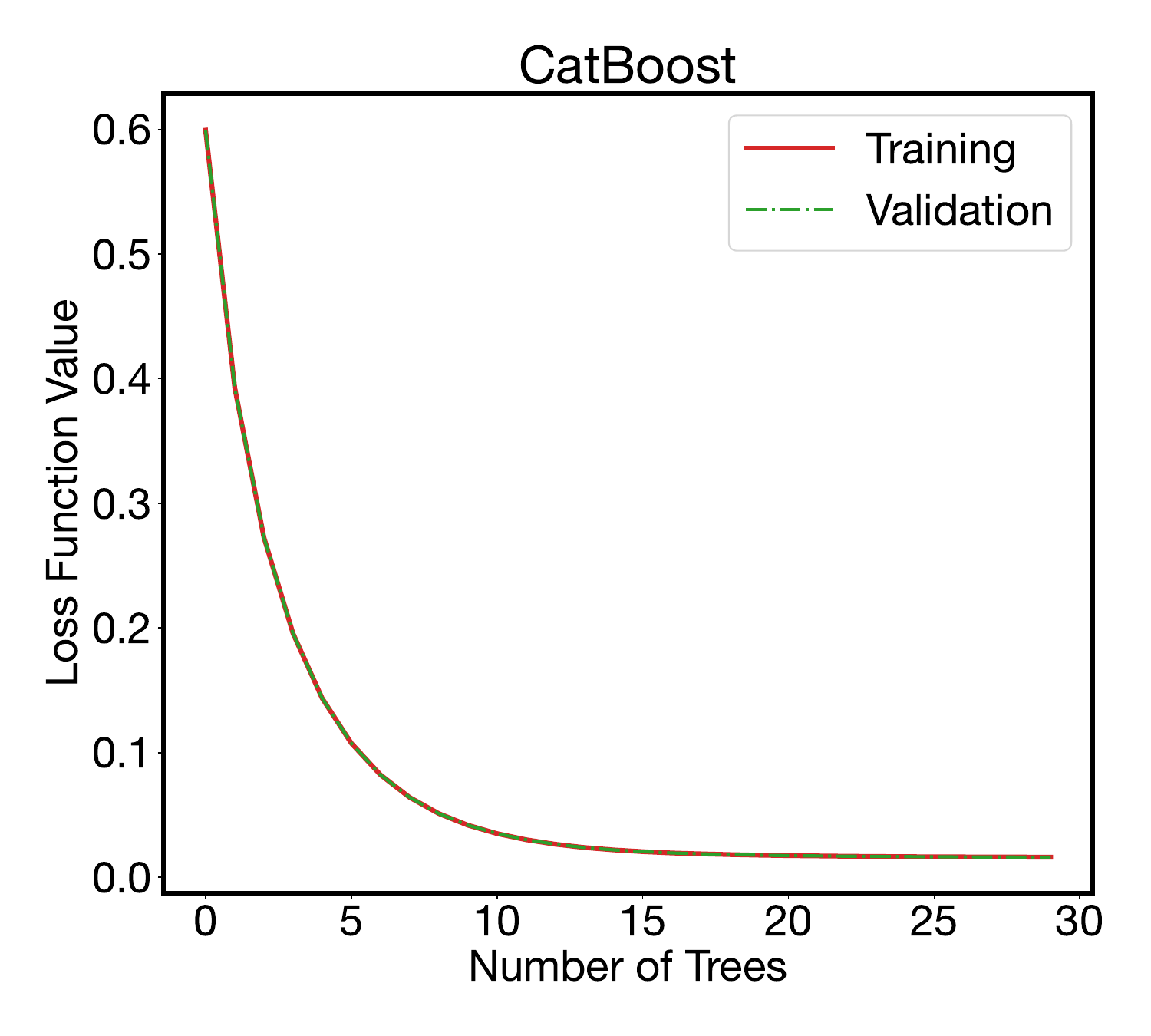}
    \includegraphics[width = 0.8\linewidth]{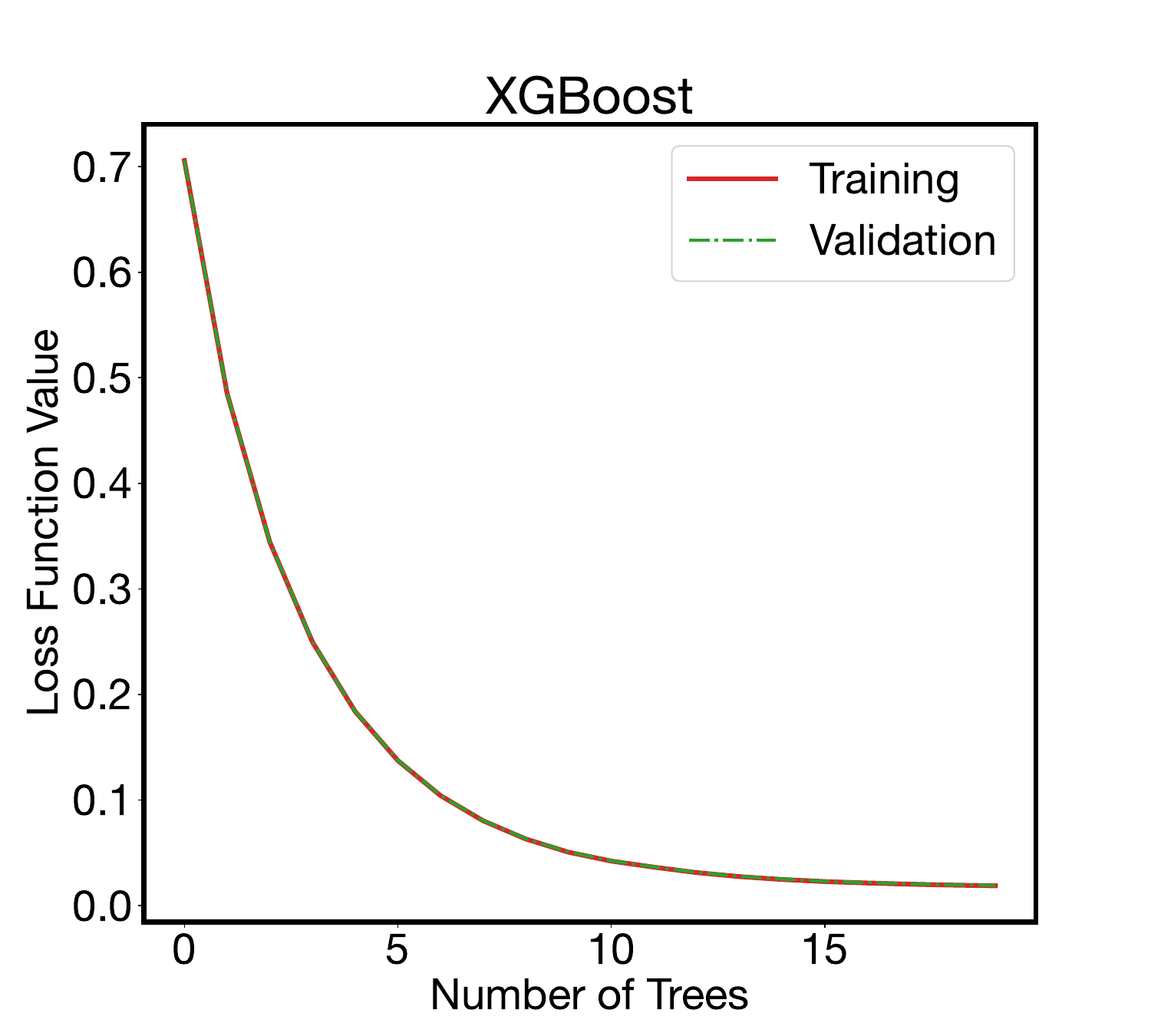}
    \caption{Learning curve of  CB and XGB.}
    \label{fig: LC}
\end{figure}

For a better understanding of the model training, we have shown the learning rate of the XGB and CB models in Fig.~\ref{fig: LC}. It serves as a pivotal tool for understanding the model’s learning trajectory and performance throughout training. The learning curves enable us to diagnose issues of underfitting or overfitting, hence ensuring the model’s robustness. Moreover, they assist in the process of hyperparameter tuning, thereby optimizing the model’s performance. Lastly, they provide insights into the efficiency of the training process, potentially conserving computational resources.

\begin{figure}
    \centering
    \includegraphics[width = 0.8\linewidth]{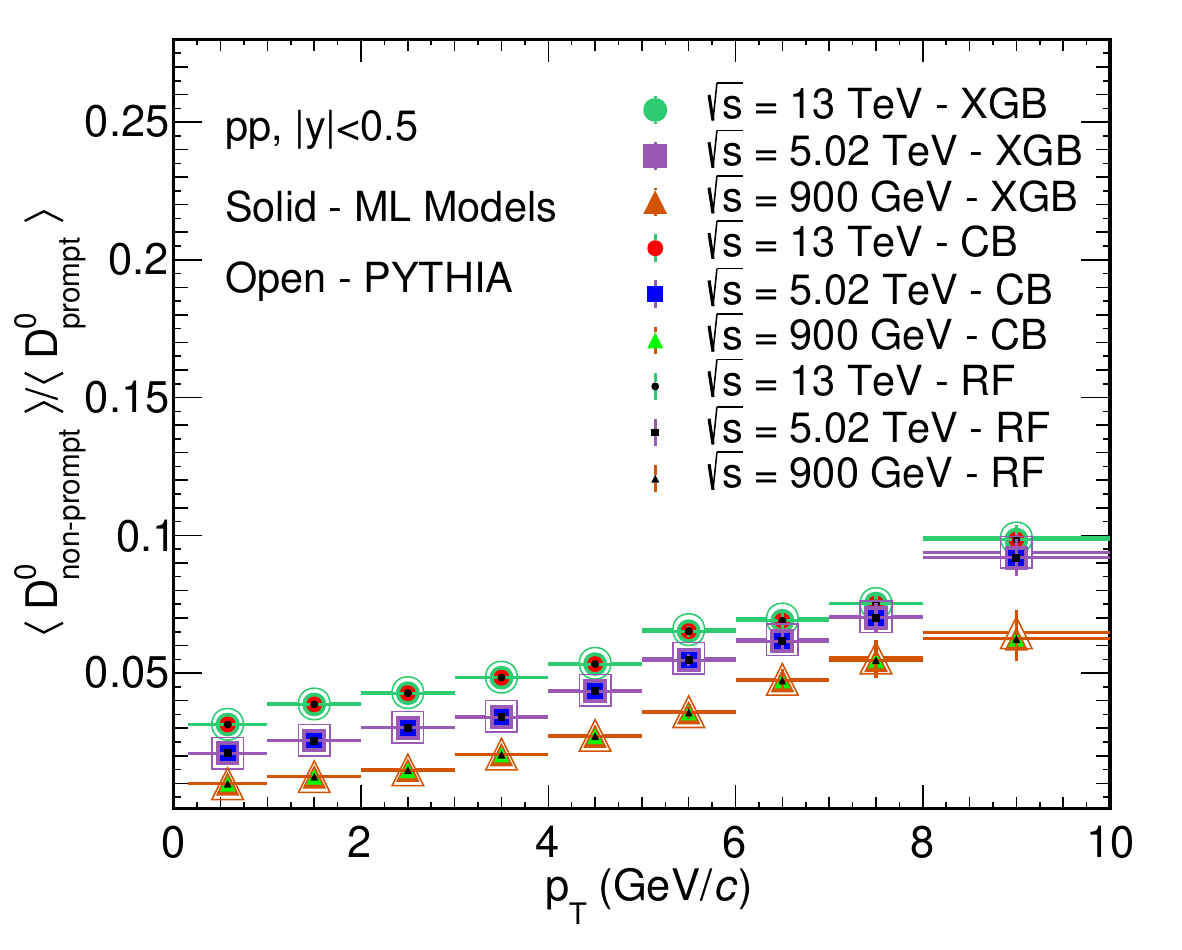}
    \caption{Nonprompt to prompt $D^0$ meson ratio for three different center-of-mass energies from PYTHIA8 compared with the predictions from XGB, CB, and RF in minimum bias $pp$ collisions.}
    \label{fig: np_p_all}
\end{figure}

\begin{figure}
    \centering
    \includegraphics[width = 0.8\linewidth]{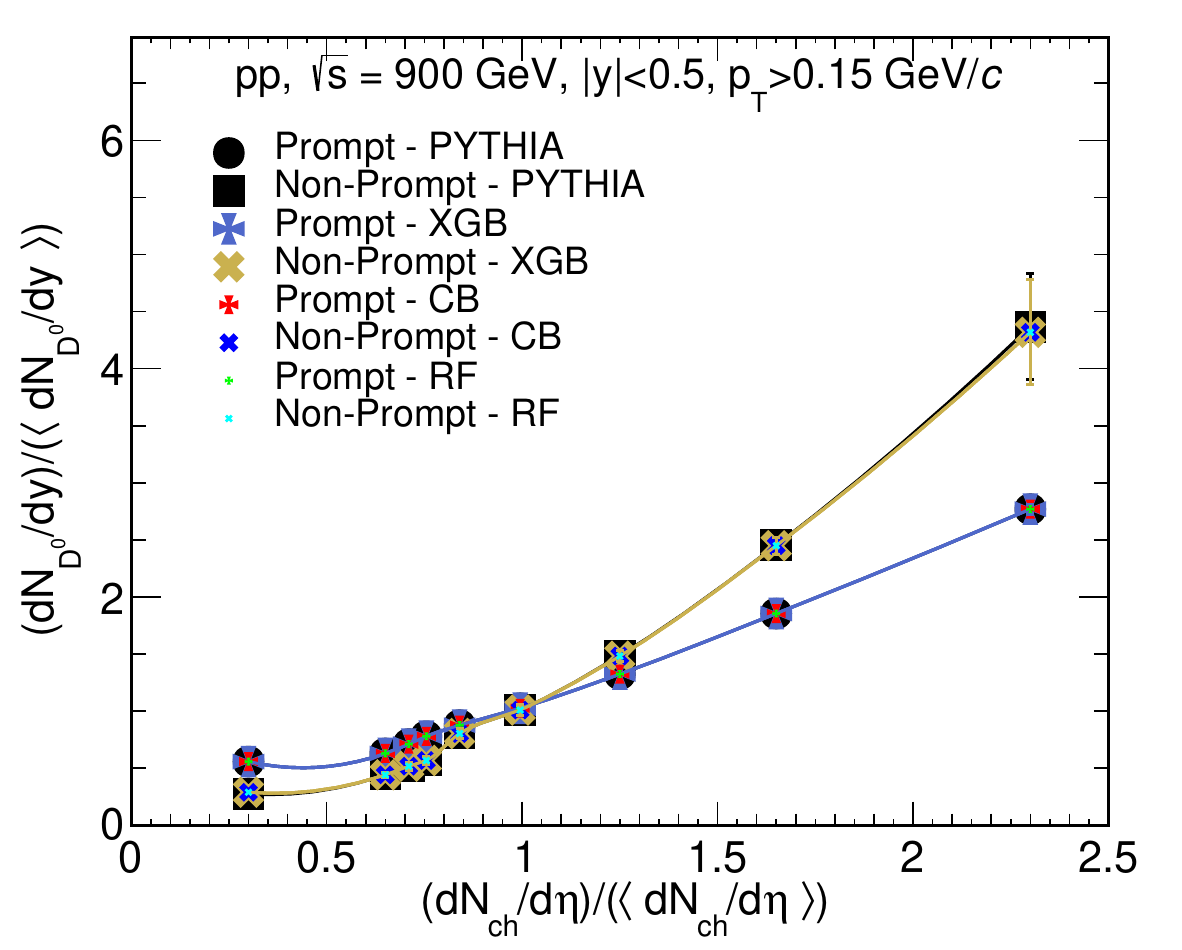}
    \includegraphics[width = 0.8\linewidth]{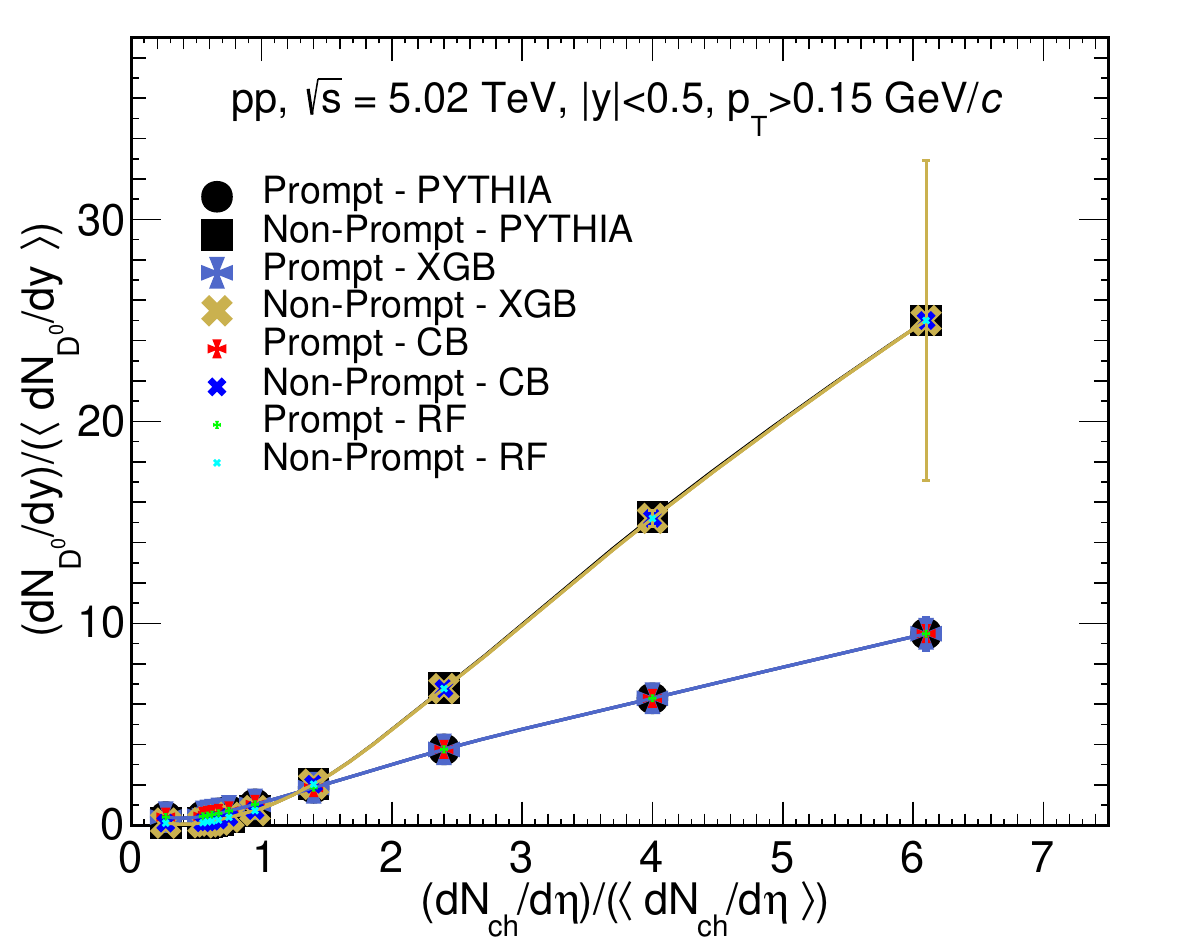}
    \includegraphics[width = 0.8\linewidth]{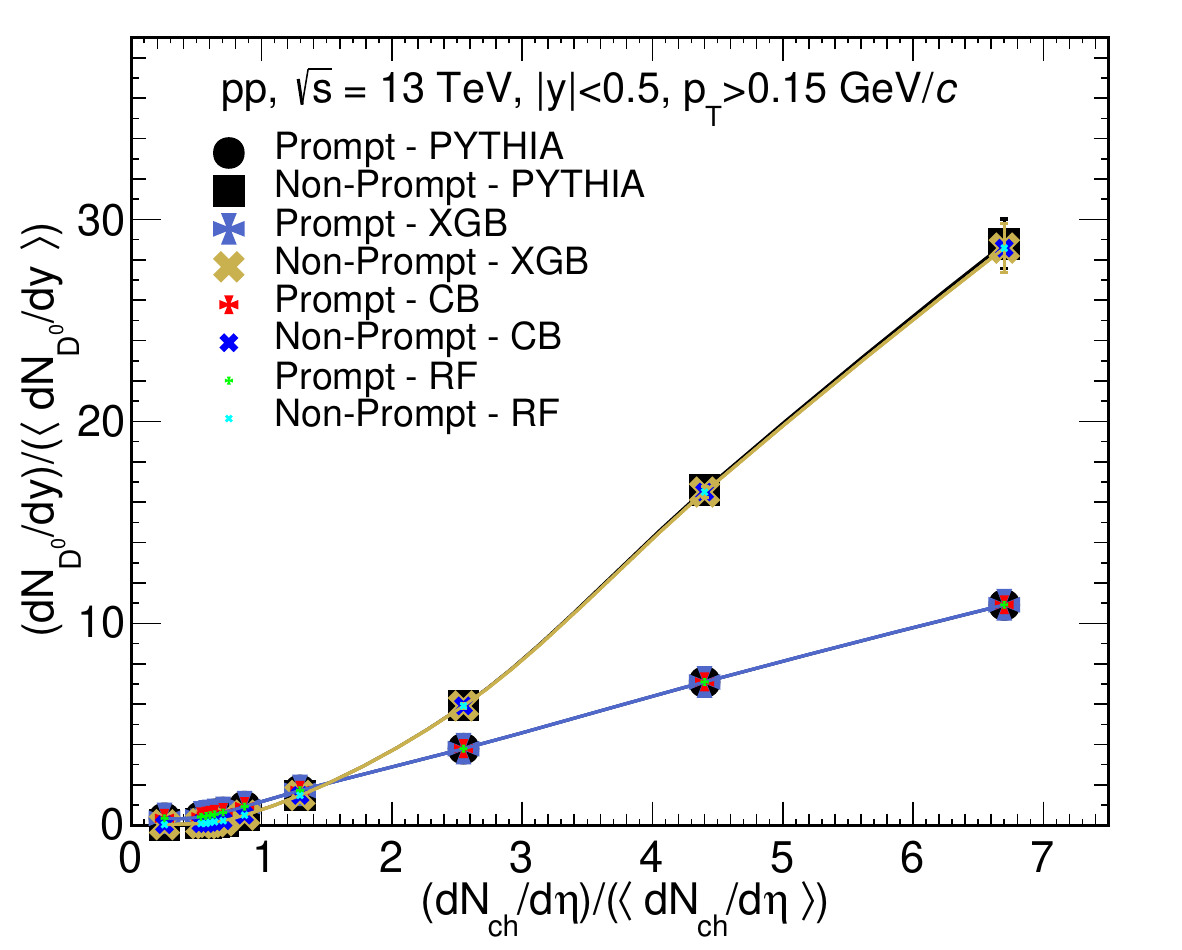}
    \caption{Self-normalized $p_{\rm{T}}$ integrated prompt and nonprompt $D^{0}$ meson yield at midrapidity ($|y|<0.5$) as a function of normalized charged-particle multiplicity in minimum bias $pp$ collisions at $\sqrt{s} = 13~\rm{TeV}$ (upper), $\sqrt{s} = 5.02~\rm{TeV}$ (middle), and $\sqrt{s} = 900~\rm{GeV}$ (lower) from PYTHIA compared with predictions from XGB, CB, and RF. The charged-particle multiplicity is obtained within the ALICE-V0 detector acceptance.}
    \label{fig: self_norm_all}
\end{figure}

\begin{figure}
    \centering
    \includegraphics[width = 0.8\linewidth]{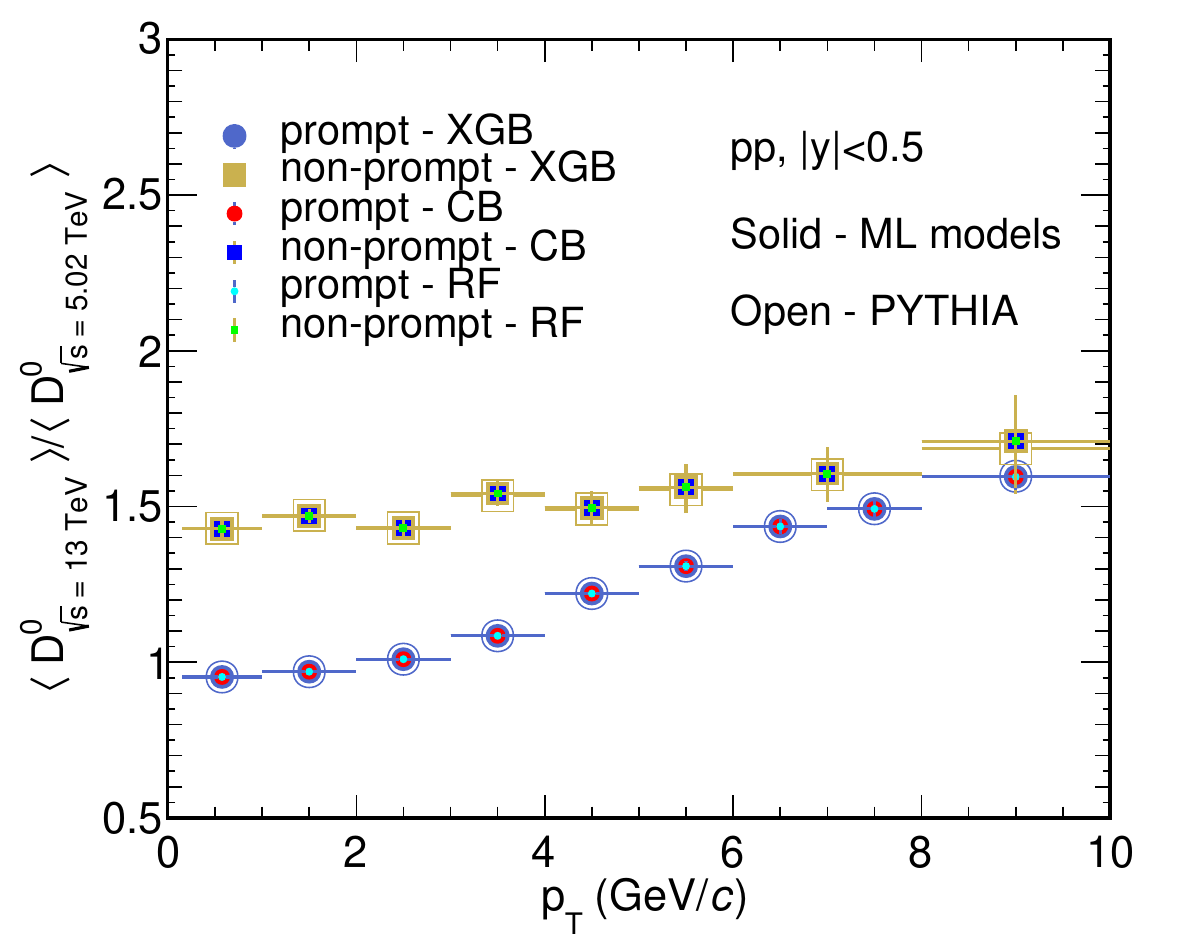}
    \includegraphics[width = 0.8\linewidth]{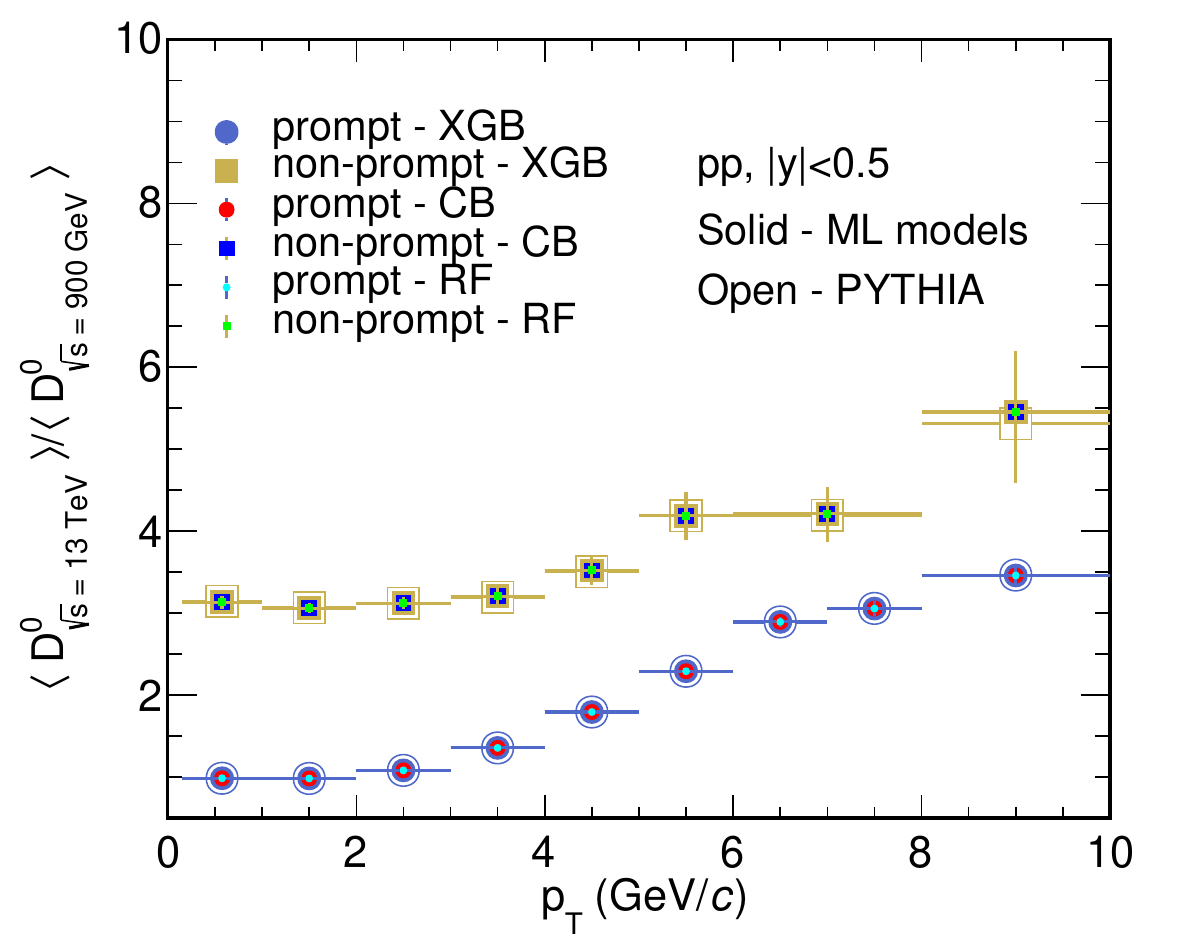}
    \caption{Ratio of $D^{0}$ yield in $pp$ collisions at $\sqrt{s} = 13$ and 5.02 TeV  (upper) and at $\sqrt{s} = 13~\rm{TeV}$ and $\sqrt{s} = 900~\rm{GeV}$ (lower) as a function of $p_{\rm{T}}$ from PYTHIA compared with the predictions from XGB, CB, and RF.}
    \label{fig: self_norm_all}
\end{figure}
However, the RF model, being an ensemble of Decision Trees, does not learn iteratively. Each tree in the forest grows independently of the others. Therefore, there is no concept of iterations during which the model progressively learns and improves. Thus, it is not possible to plot a learning curve for the RF method, unlike for the XGB and CB models. Here, one can observe that the loss function for the training and validation data sets are identical. This is because we use the $pp$ collision data simulated at PYTHIA. However, in a more realistic scenario, we might need more complex algorithms to obtain such a result.

Moreover, we replotted Figs.~\ref{fig:non-prompt_to_prompt}-\ref{fig:nrg_frac} with predictions from all the ML models: XGB, CB, and RF, for an explicit comparison, which is shown in Figs.~\ref{fig: np_p_all}-\ref{fig: self_norm_all}.

\end{document}